\def\gmode{} 

 \def\gdriver{dviout}

\newcommand{\dtitle}[1]{\title{ \if \gmode \else
\color{red} Demo mode!\\
comment out \textbackslash def \textbackslash gmode\{demo\} at the header to include figures \color{black}\\
\fi
#1 }}
\ifx\pdfoutput\undefined
\else
 \def\gdriver{}
\fi

\documentclass[\gmode,\gdriver,runningheads]{llncs}

\usepackage{graphicx}
\usepackage{amsmath,amssymb} 

\usepackage{bm}
\usepackage{siunitx}
\usepackage{wrapfig}

\usepackage{bm}
\newcommand{\fnote}[1]{}

\newcommand{\bnote}[1]{{\color{red} #1 \color{black}}}
\newcommand{\mnote}[1]{}
\newcommand{\kcut}[1]{} 

\newcommand{\scut}[1]{}
\ifx\pdfoutput\undefined
\else
 \usepackage{CJKutf8}
 \renewcommand{\fnote}[1]{}
 \renewcommand{\mnote}[1]{}
 \renewcommand{\bnote}[1]{}
\fi

\newcommand{\ie}{{\it i.e.}}

\newcommand{\um}[1]{{\SI{#1}{\micro \metre}}}

\newcommand{\figref}[1]{{Fig.\ref{fig:#1}}}

\newcommand{\equref}[1]{Eq.(\ref{equ:#1})}
\newcommand{\secref}[1]{Sec.\ref{sec:#1}}

\graphicspath{{../figs/}{./figs/}}

\begin{document}
\pagestyle{headings}
\mainmatter
\def\ACCV20SubNumber{719}  


\title{Dense Pixel-wise Micro-motion Estimation of Object Surface by using
Low Dimensional Embedding of Laser Speckle Pattern
  }

  \ifx
Recovery of Micro-Movements of Object Surface

Spatio-temporal Analysis
Laser Speckle
Embed into low dimension

Recovery of Micro-Movements of Object Surface
by Embedding Spatio-temporal variation of Laser Speckle pattern into 
low dimension space

Embedding Spatio-temporal Information of Laser Speckle pattern 
into low dimension space
for Micro-Movements Recovery of Object Surface
\fi


\titlerunning{Dense Pixel-wise Micro-motion Estimation of Object Surface}
%
\author{Ryusuke Sagawa\inst{1}\orcidID{0000-0002-6778-8838} \and
Yusuke Higuchi\inst{2} \and \\
Hiroshi Kawasaki\inst{2}\orcidID{0000-0001-5825-6066} \and
Ryo Furukawa\inst{3}\orcidID{0000-0002-2063-1008} \and
Takahiro Ito\inst{1}\orcidID{0000-0003-2886-1067}
}
\authorrunning{R. Sagawa et al.}
%
\institute{National Institute of Advanced Industrial Science and Technology \and
Kyushu University \and
Hiroshima City University
}


\maketitle

\begin{abstract}
  This paper proposes a method of estimating micro-motion of an object
  at each pixel
  that is too small to detect under a common setup of camera and illumination.
  The method introduces an active-lighting approach to make the
  motion visually detectable. The approach is based on speckle
  pattern, which is produced by the mutual interference of laser light
  on object's surface and continuously changes its appearance according to the out-of-plane 
  motion of the surface.  
  In addition, speckle pattern becomes
  uncorrelated with large motion. To compensate such micro- and large motion, 
the method estimates the motion
  parameters up to scale at each pixel by nonlinear embedding of the speckle pattern into
  low-dimensional space. The out-of-plane motion is calculated by making the
  motion parameters spatially consistent across the image. In the experiments,
  the proposed method is compared with other measuring devices to prove the
  effectiveness of the method. 
\end{abstract}

\section{Introduction}

The analysis of physical minute movements of objects is important to
understand the behavior of various systems including mechanical
systems, fluids analysis and biological studies. For example,
mechanical systems vibrate at various frequency
distributed over a wide range from several hertz to
several kilohertz only with the extremely small amplitude, such as less than one
millimeter. There are many approaches to measure such a minute movements, and
typical methods are to use vibration sensors based on various
measurement principles including accelerometers and displacement gauges.
Although these methods can measure such a minute movements precisely, there is
a severe limitation, such as only a single point can be measured by each sensor,  
because of necessity of physical contact, resulting in a measurement of only limited and 
small areas. 
In order to conduct accurate and robust analysis on spatio and temporal effect 
of vibration, measurement of micro movement of dense and  wide region is required,
which cannot be achieved by existing sensors and methods.

\ifx
Visualization of the movements of a precision machine is
one of the important applications. If a machine has a movable part
such as a machine tool, suppression of its vibration is indispensable
for improving the precision and such vibration is usually measured by
contact sensors. In order to investigate how the vibration is
propagated, dense measurement is required, whereas contact sensor can
only measure sparse sample points. Note that, as the first step of the
investigation, the magnitude of the vibration is not necessarily
important, but the density and frame rate are important.
\fi


Acquiring such wide area's micro-motion like vibration of almost rigid but deformable object 
surfaces requires dense observation in time and space.
One promising solution is to use a camera, since it can capture wide area 
information at a short period of time.
Optical triangulation is a common method to measure displacement from captured 
image, 
however, since the movement by 
vibration is much smaller than the size of the object, \ie, less than one pixel, 
it is impossible to measure such minute motion by the method.
Another approach is to use optical
interferometry. Since the method uses the difference of the length of
multiple light paths, the accuracy mainly depends on the wavelength of
the light, which indicates that the resolution of a camera is less
dependent. The disadvantage of this approach is that it
requires a complicated and precise setup with many optical components
to generate the high quality reference light and path.

\begin{figure}[t]
  \begin{center}
    \begin{tabular}{ccc}
    \includegraphics[width=0.38\textwidth]{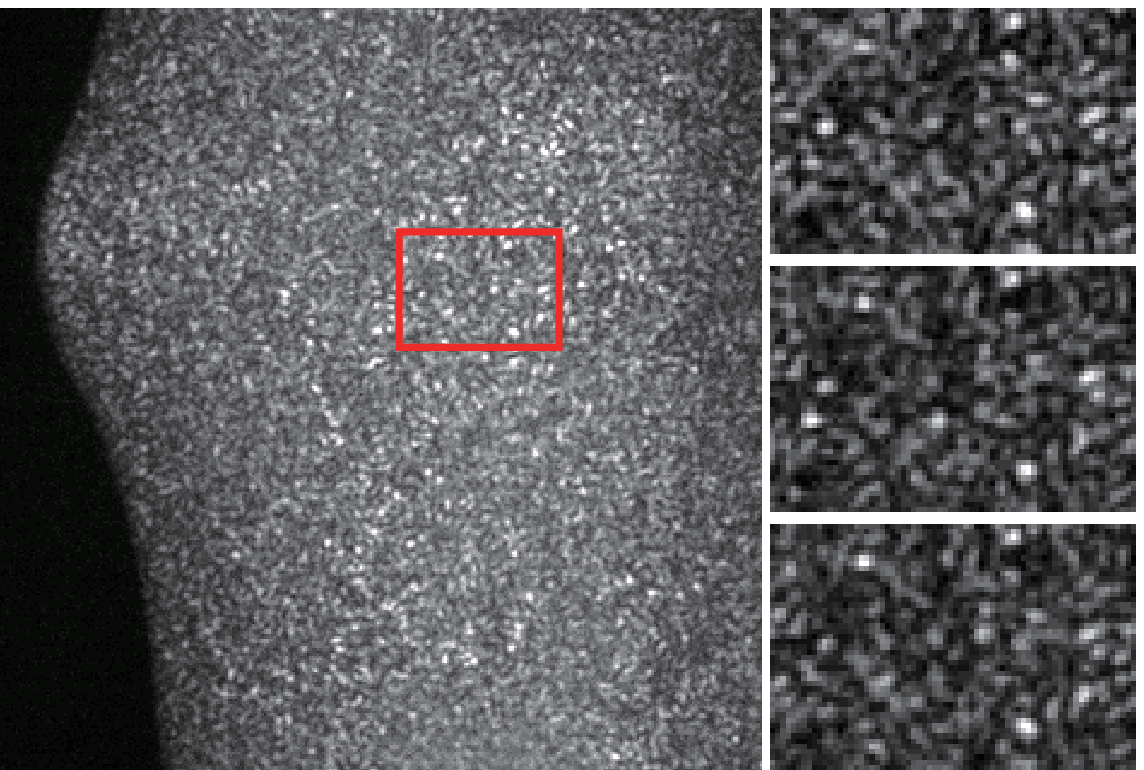} &
    \includegraphics[width=0.3\textwidth]{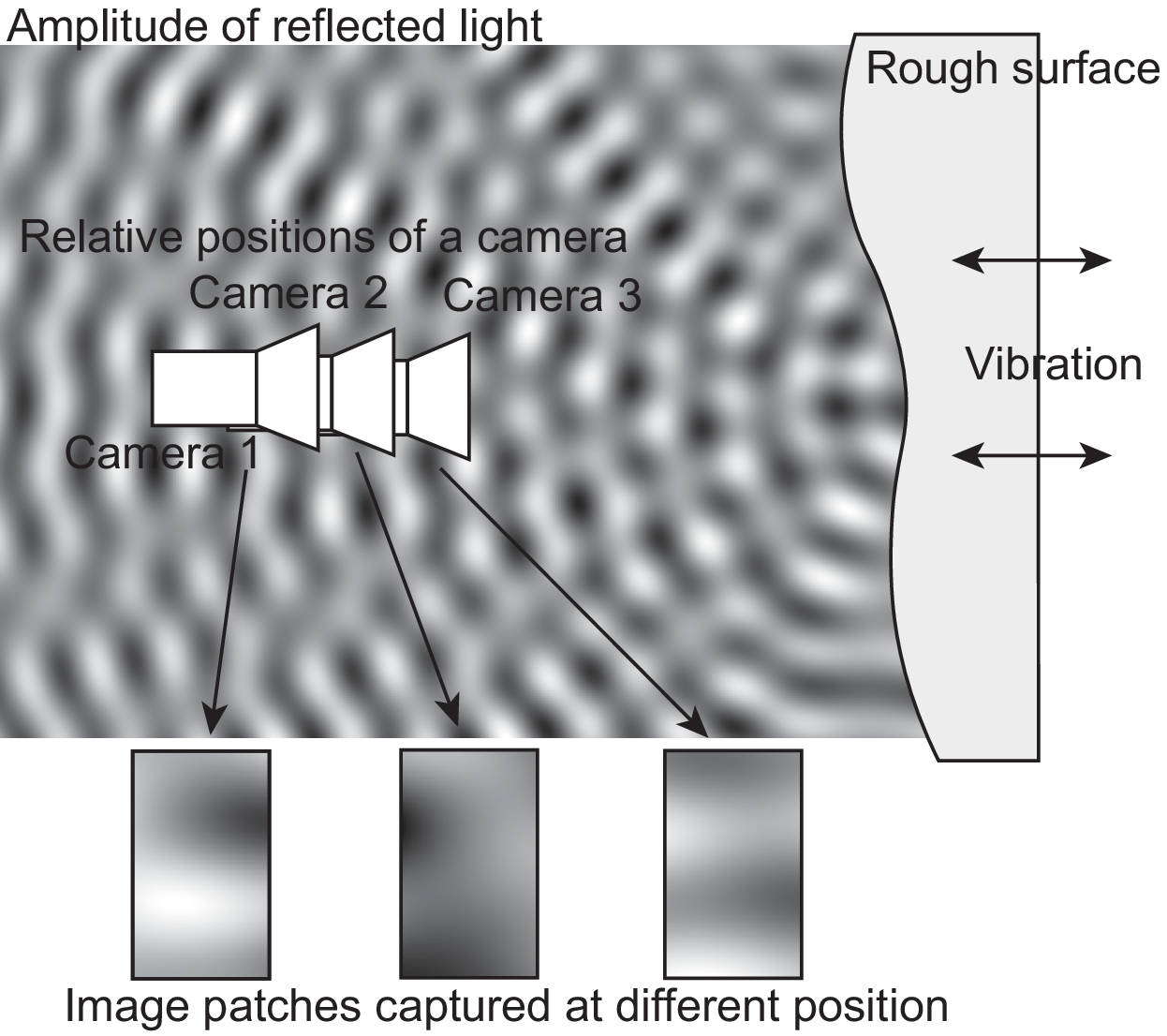} &
    \includegraphics[width=0.28\textwidth]{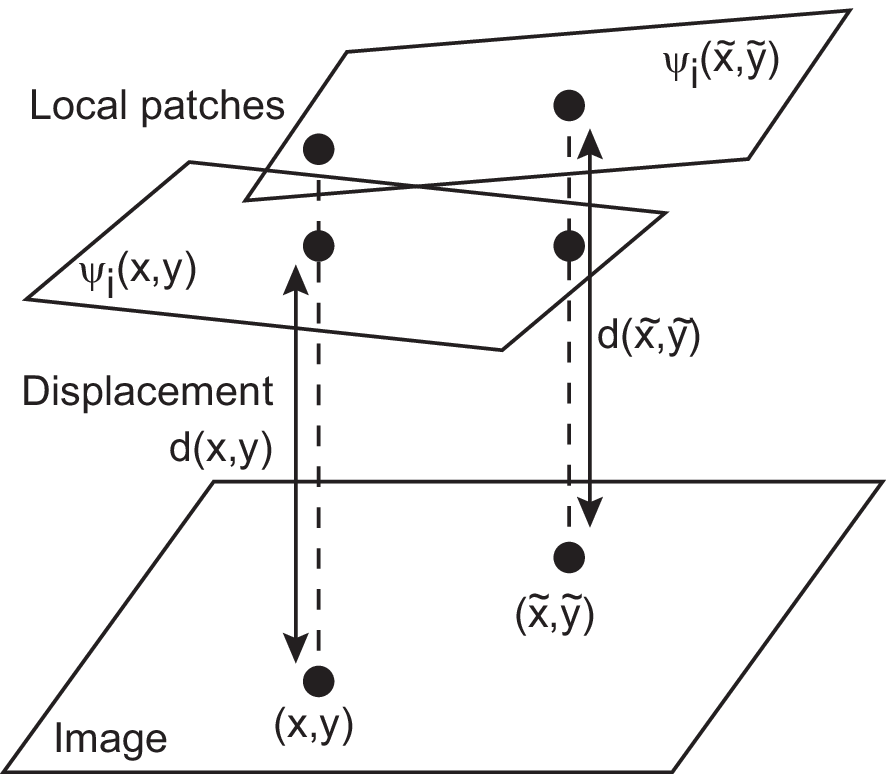} \\
    (a) & (b) & (c) \\
    \end{tabular}
  \end{center}
  \caption{An example of speckle pattern and modeling motion of piecewise planes. (a) An example of speckle
    pattern on an object's surface is shown. While the power of the incident
    light is uniform, the dotted pattern is observed by a camera. (b) The
    amplitude of light changes drastically according to the position due to the
    mutual interference. (c) The plane parameters $\bm{\psi}_i(x,y)$ and
    $\bm{\psi}_i(\tilde{x},\tilde{y})$ are to be estimated.}
  \label{fig:speckle_example_modeling}
\end{figure}

Speckle pattern, which is a phenomenon of light interference on object surface, 
has also been frequently used for detecting micro-motion.
Since one of important characteristics of speckle pattern is 
high sensitivity of its appearance, it is utilized for observing minute 
phenomenon, such as micro-motion of object surface.
Another unique characteristic of speckle pattern is that no reference light is 
necessary to observe the pattern, because it is mutual interference of the
lights reflected from the multiple points on object's surface. Those characteristics
enables a simple setup of devices to detect micro-motion in a
wide field of view. 
%
%
A typical usage of speckle pattern for observing the minute movement is the 
frequency analysis, since the minute movement of object surface caused by
vibration usually has a cyclic period, which can be effectively calculated by 
temporal analysis of the intensity at each pixel.
One limitation of such speckle analyses is that since speckle is 
essentially an independent phenomenon at each pixel, it cannot extract temporal 
relationship, such as phase,
between neighboring pixels, \ie, wave propagation on object surface caused by vibration 
cannot be acquired. To the best of our knowledge, such  wave propagation phenomenon 
has not been measured yet.
%
In this paper, we propose a method to
extract such spatio-temporal information of micro-motion on the surface by 
analyzing speckle pattern captured by a camera. In our method, series of 
speckle patterns of each patch are embedded into 
low-dimensional space and surface parameters at each pixel are estimated by 
joint optimization.
The contribution of the paper is summarized as follows.
\begin{itemize}
  \item The minute movement at each point of object's surface can be extracted
        as the low-dimensional embedding of laser speckle pattern. 
  \item The spatial consistency of the embedded information is
        globally supported by smoothness constraint of the surface,
        which achieves the spatio-temporal analysis of speckle patterns.
  \item Various kind of wave propagation phenomena caused by 
	vibration are first densely measured and visualized by our algorithm in 
	real scene experiments. 
\end{itemize}

\section{Related work}

As methods to observe minute movement of an object, four methods can
be considered. First, a contact sensor such as accelerometers is
typically used to measure the vibration, but this approach is not
suitable for spatial analysis that needs dense observation.
Non-contact methods that observe vibration by using cameras
have advantage in dense observation.

Both passive and active approaches have been studied to observe minute
movements of objects by using a camera. As passive approaches, the
methods based on optical flows and pattern matching by correlation
have been proposed to observe vibrating
objects~\cite{demilia13:_uncer,caetano11:_vision}. Wu et
al.~\cite{Wu12Eulerian} proposed a method to visualize subtle
movements by magnifying the displacements based on spatio-temporal
filtering. They have extended the method to recognize the sound by
observing the vibration in the video~\cite{Davis2014VisualMic}
or the method to estimate material properties~\cite{davis17:_visual_vibrom}. 
If the displacement is small, these methods need to zoom up the targets to detect the change of intensity
due to their passive approach. Since passive approaches cannot be
applied in the case that the vibration is smaller than the camera
resolution, they have disadvantage in sensitivity.



As active approaches, laser-based methods have been proposed.
When the wide field of view is illuminated by coherent laser light,
\figref{speckle_example_modeling}(a) is an example of speckle pattern on an
object's surface. While the power of the incident light is uniform,
the dotted pattern is observed by a camera.  When the surface
vibrates, the pattern randomly changes as shown in the right three
images, which are the zoom-up of the image patch in the red rectangle.
The size of speckle depends on the aperture of the imaging system. If
the aperture is large, the speckle of neighboring pixels is unrelated,
and hard to distinguish from the noise of the imaging system in some
cases. The irradiance of speckle pattern obeys negative exponential
statistics, which indicates that some bright points exist while
speckle patterns generally have many dark areas. The basic
characteristics of speckle pattern have been studied in
literature~\cite{goodman76:_some,dainty84:_laser_speck}.

There are two approaches to extract the information from the images of
speckle. First one that calculate the phase of vibration by using
speckle pattern is based on interferometry, which is called electronic
speckle pattern interferometry
(ESPI)~\cite{loekberg76:_vibrat,creath85:_phase,bavigadda12:_vibrat}.
This approach uses a reference light and compare the length of the
paths between the reference light and the light from an object. If the
length changes by the deformation of the object, the intensity
captured by a camera changes due to the interference of two lights.
In this case, the object movement is called out-of-plane motion that
is along the viewing direction of the camera. The absolute
displacement by the movement can be calculated by using the wavelength
of the light. But since this approach requires a highly calibrated
setting, it have disadvantage in simplicity of the data acquisition.
Garc{\'\i}a et al.~\cite{garcia2008three} measured distance between object and
camera by out-of-plane motions, which are not based on interferometry but
needs precise calibration to obtain the speckle beforehand that corresponds 
to each distance.

\begin{figure}[t]
  \tabcolsep = 2mm
    \begin{center}
  \begin{tabular}{cc}
      \includegraphics[height=0.2\textwidth]{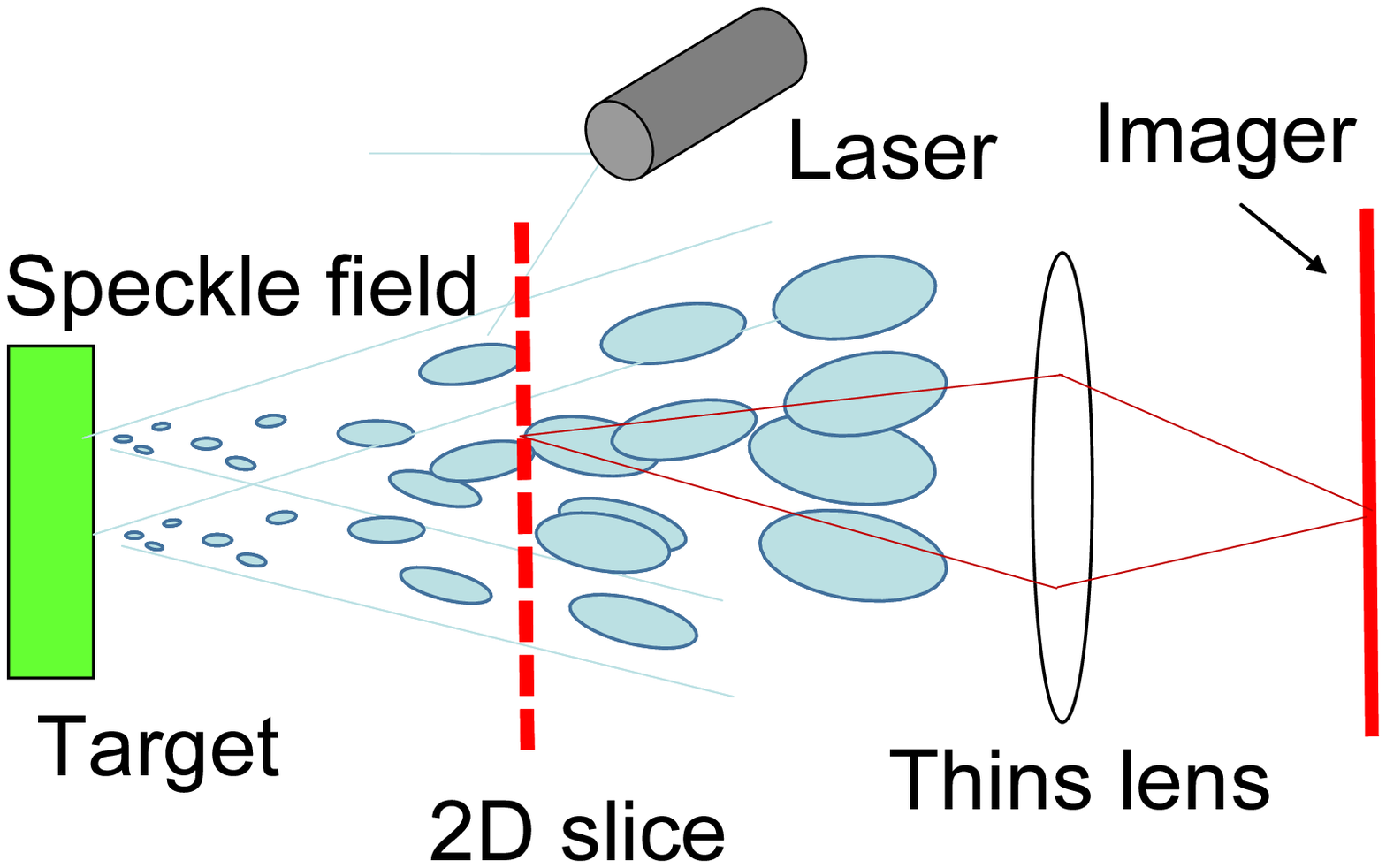}&
      \includegraphics[height=0.2\textwidth]{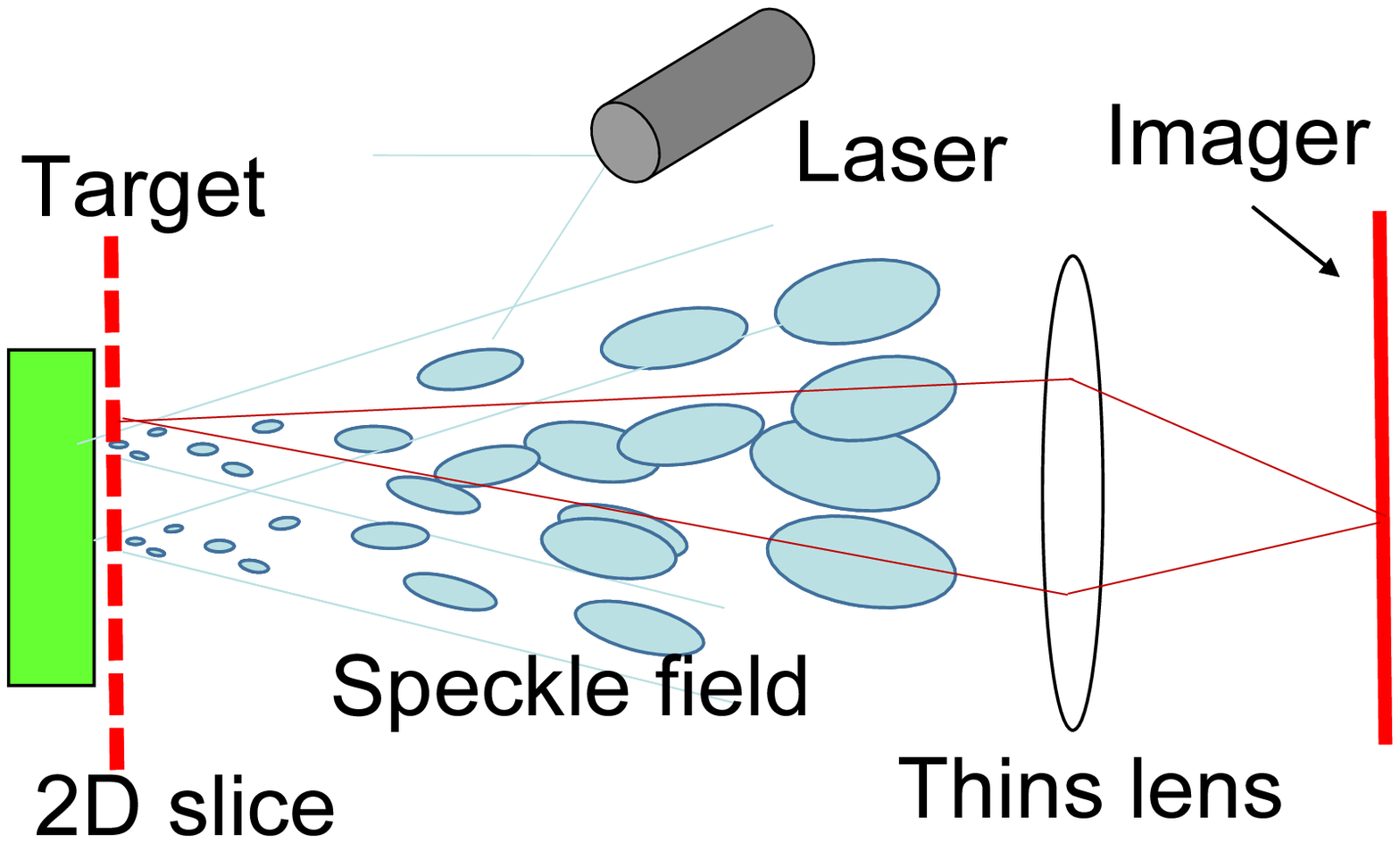}\\
(a)&(b)\\
  \end{tabular}
\end{center}
    \caption{The speckle field and observed 2D slices. (a) thin lens observation
    of fused speckle and (b) thin lens observation of independent speckle
    pattern.}
      \label{fig:speckle_field}
\end{figure}

Second approach with speckle pattern, which is called speckle
photography, do not require the setup for interference.  If an object
slightly moves in the direction parallel to the image plane, which is
called in-plane motion, the movement of speckle pattern is simple
translation~\cite{yamaguchi80:_real_time}.  The motion of the object
therefore can be calculated by the techniques of pattern matching.
This approaches are used for estimating blood
flow~\cite{fujii87:_evaluat_part}, motion sensing
~\cite{zizka11:_speck}, detecting small
deformation~\cite{chang12:_laser_speck}. 
For the analysis of out-of-plane motion, Zalevsky et
al.~\cite{zalevsky09:_simul} proposed a method to extract the
vibration caused by sound and heart beat by using speckle patterns
produced by projecting a point laser. Although the periodic signal is
detected by calculating the correlation of the pattern, it is
single-point observation and cannot applied to spatial analysis of the
vibration. 
Synnergren \cite{synnergren1997measurement} proposed to combine stereo vision 
and speckle photography, 
where measuring in-plane motions using speckle photography 
is used for estimate disparity between stereo images. 
Thus, out-of-plane displacements can be estimated. 
The speckle motion of a front-parallel object along out-of-plane axis becomes 
rotationally symmetric as analyzed in \cite{jakobsen2012spatial}, which can be used
to estimate out-of-plane motion.
The methods~\cite{jacquot1979speckle,jo2015spedo,smith2017colux,smith2018tracking} 
to estimate combined motion including in-plane and out-of-plane motions are also proposed.
Although these methods are based on tracking speckle patterns,
one of the problem is that a speckle pattern drastically change along 
out-of-plane motion if it is observed at the focal plane.
They therefore used defocused or lens-less setups to observe moderate change of speckle.
As the trade-off, the spatial resolution of the image is degraded and
only a couple of motions are estimated at a time.
In contrast, the method proposed in this paper 
observes speckles on the target surface using lens.
It can obtain pixel-wise displacement information as a result,
which is realized by multivariate analysis by using speckle pattern 
for spatial analysis of out-of-plane vibration instead of tracking pattern.

\section{Overview of the method}

\subsection{Principle}


Generally, previous speckle photography methods observe 2D slices of 3D speckle
field that is far from the target surface, which is generated by interference of
reflected laser, as shown in \figref{speckle_field}(a). Since the speckle fields are magnified
and fused, changes of the speckle pattern is analyzed as a whole. However, this
makes the spatial resolution inevitably low, or large rigid motions of objects are
assumed. 
%
One solution is to observe 2D slices near the 
object surface, 
where each speckle pattern is small and independent,
by focusing on the target surface 
(\figref{speckle_field}(b)). 

\ifx
\begin{wrapfigure}{r}[10pt]{0.4\textwidth}
  \centering
      \includegraphics[height=0.2\textwidth]{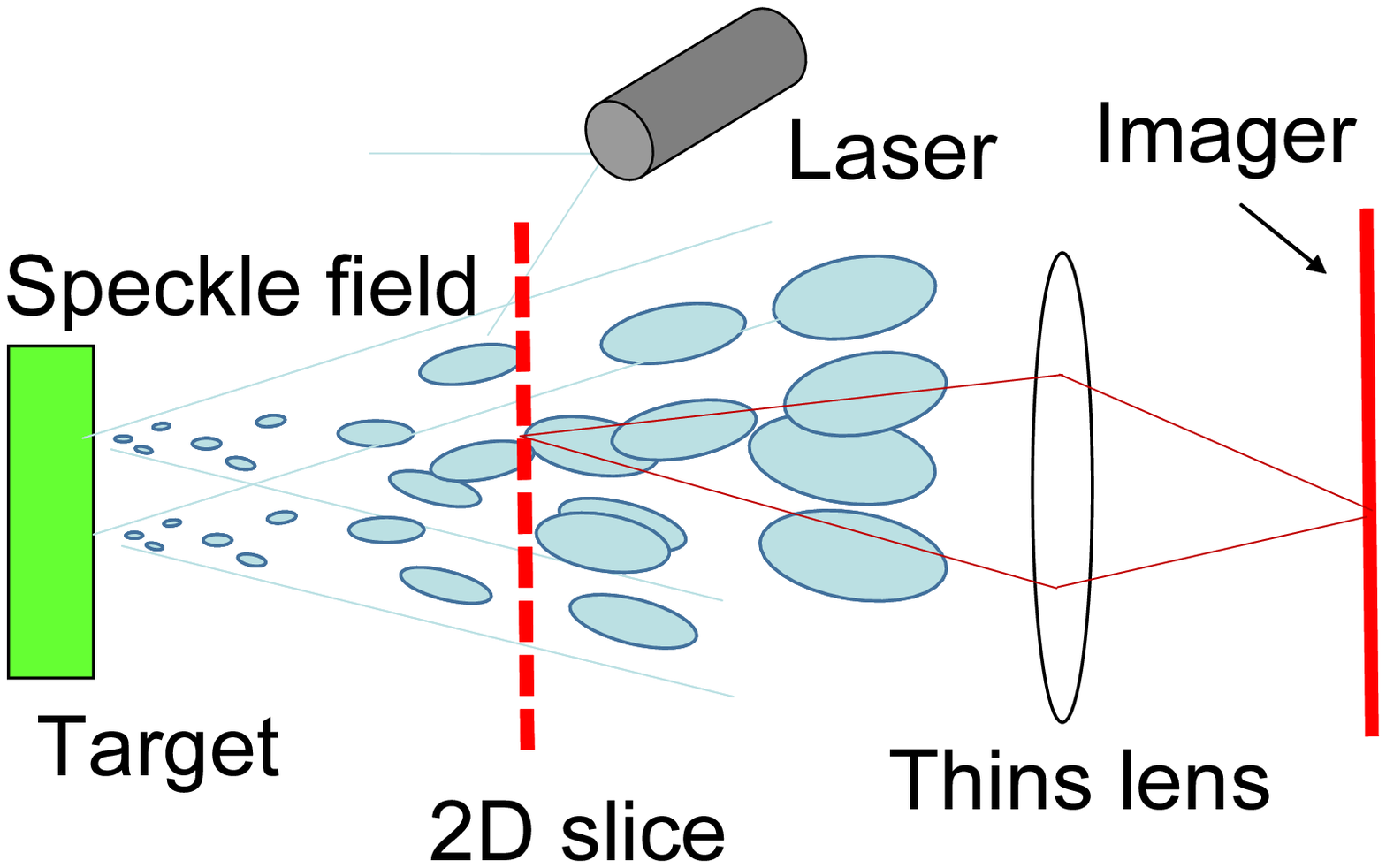}\\
      \includegraphics[height=0.2\textwidth]{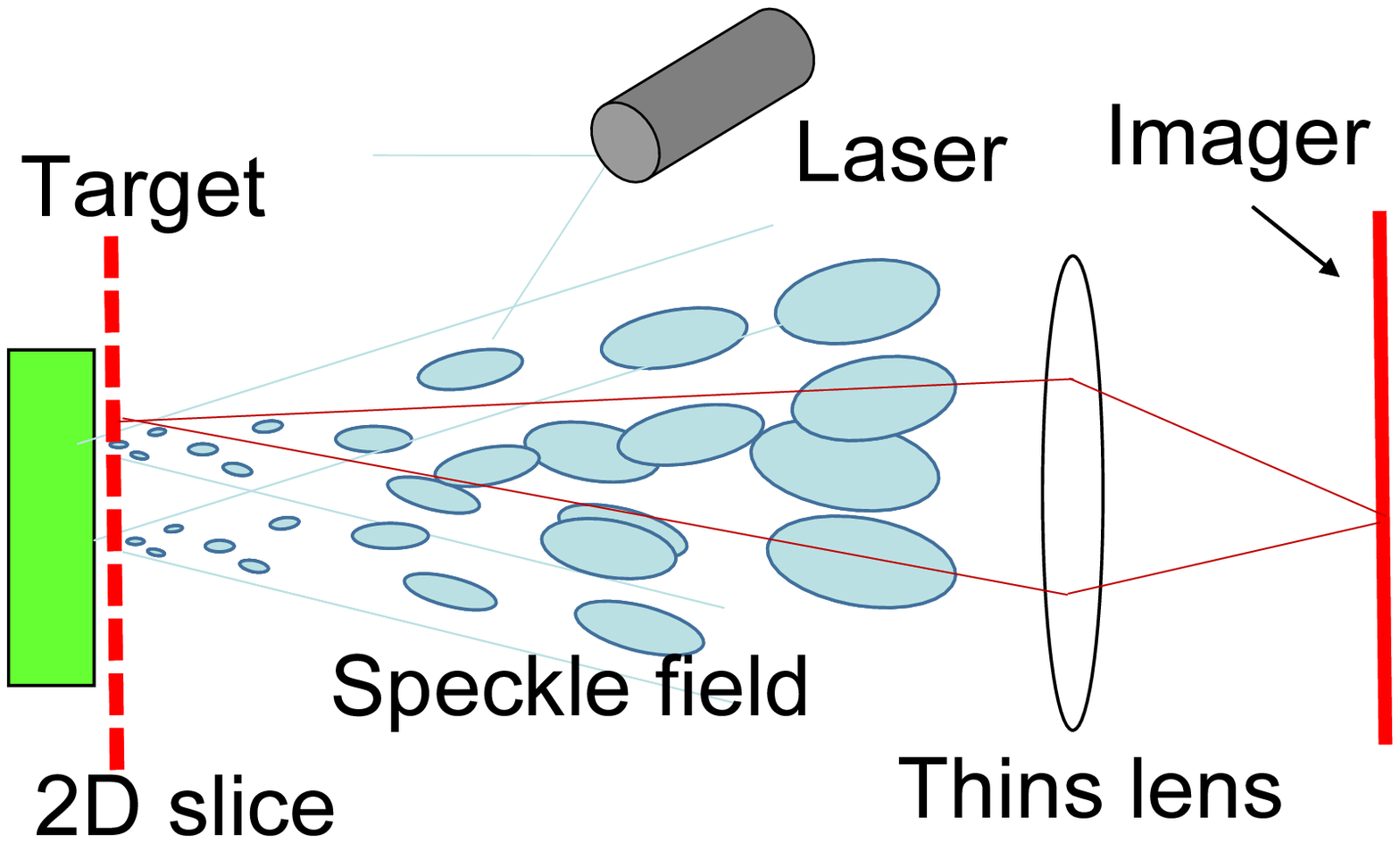}
    \caption{The speckle field and observed 2D slices. (top) thin lens observation
    of fused speckle and (bottom) 
independent speckle pattern.}
      \label{fig:speckle_field}
\end{wrapfigure}
\fi
%
However, there are two problems in this approach: 
(1) the speckle pattern near object surface drastically changes its appearance
w.r.t. out-of-plane motion,
and 
(2) if focus plane is near the target, 
speckle image w.r.t. out-of-plane motion becomes totally different~\cite{jakobsen2012spatial},
because 
the speckle fields from local points are not fused. 
Thus, analysis of this type of speckle images 
needs a completely different approach.
To overcome the problem, the intensities in a local image patch around a pixel is considered
as a feature vector of the pixel and the distance between feature vectors
can be used as a metric of the similarity/difference between the slices.

\ifx
The key idea of the proposed method is that the speckle pattern is
suitable to observe minute movements of an
object, since it is sensitive to the relative position between the object and
a camera. 
\figref{speckle_example_modeling}(b) shows a situation that a coherent light is
reflected on multiple points on the rough surface of an object. Since
the reflected lights have different phases, the mutual interference
occurs between them. When the relative position between the surface
and camera changes along the out-of-plane axis, the observed image
patches change drastically according to the position of the camera
relative to the object surface. Although the amplitude of speckle
pattern is almost random if the number of reflected points is very
large, the image patches are reproducible if the relative position is
the same.

If the intensities in a local image patch around a pixel is considered
as a feature vector of the pixel, the distance between feature vectors
can be used as a measure of the difference between the surface poses.
However, since the speckle pattern is random, the comparison 
of local patches is meaningful for the same point of the surface
if in-plane motion is not considered.
Therefore, we assume that the motion of surface is 
out-of-plane motion along the line of sight
in this paper, and compare local patches at the same pixel
to analyse the minute movement of the surface.
\fi

\subsection{System configuration and algorithm}

\figref{speckle_example_modeling}(b) shows a configuration of the system.
A coherent laser light is
reflected on multiple points on the rough surface of an object and since
the reflected lights have different phases, the mutual interference
occurs between them, generating speckle field. When the relative position between the surface
and camera changes along the out-of-plane axis, the observed image
patches change drastically according to the position of the camera
relative to the object surface. Although the amplitude of speckle
pattern is almost random if the number of reflected points is very
large, the image patches are reproducible if the relative position is
the same.
The proposed method consists of the following steps.
\begin{enumerate}
  \item Capture sequence of speckle pattern near object surface using wide aperture lens.
  \item Embed the feature vector of the local patch around each pixel into nonlinear 
	low-dimensional space.
  \item Make the low-dimensional space consistent between neighboring pixels.
  \item Optimize the parameters of local surface to fit the low-dimensional space.
\end{enumerate}

\section{Implementation}

\subsection{Representation of object surface}

The purpose of our method is to estimate subtle motion of object surface with 
pixel-wise density. 
The surface is assumed locally planer and 
the proposed method estimates 
3D plane parameters around each pixel $(x,y)$ 
at frame $i$ defined as follows:
\begin{align}
  z = [\tilde{x} - x, \tilde{y} - y, 1] \bm{\psi}_i(x,y)
\end{align}
where $\bm{\psi}_i(x,y) = (\psi_1,\psi_2,\psi_3)^T$ is 
three dimensional vector as the plane parameters, and
$(\tilde{x},\tilde{y})$ are neighbor pixels around $(x,y)$.
The offset at the the origin of the local plane
indicates surface displacement $d(x,y)$ as the movement at the pixel $(x,y)$,
which is calculated by $d(x,y) = [0,0,1] \bm{\psi}_i(x,y) = \psi_3$.
The local plane is calculated for each pixel, and
\figref{speckle_example_modeling}(c) shows a situation that 
the local patch of the plane for $(x,y)$ and $(\tilde{x},\tilde{y})$.
Since the displacement is estimated from image patches, it is determined up to scale.

\subsection{Embedding speckle pattern for each pixel}
\label{sec:embedding}

To observe the movement of surface, a video is captured during 
the movement with projecting the coherent light generated by a laser light source.
Let $I_i(x,y)$ the intensity of a pixel $(x,y)$ at frame $i$ of the video.
The pixels $(x_k,y_k)$ in the local image patch $P(x,y)$ forms
a feature vector $\bm{F}_i(x,y) = (I_i(x_1,y_1),\ldots,I_i(x_k,y_k))$ 
that describes the state of the local patch of surface.
To analyze the movement between two frames $i$ and $j$,
the difference of two image patches is calculated by the Euclidean distance
of the feature vectors.
\begin{align}
  D_{i,j}(x,y)^2 & = \parallel \bm{F}_i(x,y) - \bm{F}_j(x,y) \parallel^2 \label{equ:distance} \\
                 & = \sum_{(x_k,y_k) \in P(x,y)} (I_i(x_k,y_k) - I_j(x_k,y_k))^2
\end{align}
If the size of local patch is sufficiently large,
the degree of freedom of the minute movement of the surface
is much smaller than the dimension of $\bm{F}_i(x,y)$.
It indicates that the distance is approximated by the distance
of the low-dimensional vectors.
Let $\bm{\Psi}_i(x,y) = (\psi_{i,1}(x,y), \ldots, \psi_{i,l}(x,y))$
a low-dimensional vector of which the dimension is $l (<< k)$.
Namely, the distance becomes
\begin{align}
  D_{i,j}(x,y)^2 = \parallel \bm{\Psi}_i(x,y) - \bm{\Psi}_j(x,y) \parallel^2
\end{align}
The low-dimensional vector can be obtained by the techniques of
dimensionality reduction, which embed the input vectors
in the low-dimensional space so that the distance between
two vectors are preserved after embedding. Since the movement is large, 
the speckle patterns between two frames becomes almost uncorrelated.
Therefore, the relationship between the movement and the distance
of speckle patterns is expected to be nonlinear.
Various methods of dimensionality reduction for nonlinear distances
have been proposed such as Isomap~\cite{tenenbaum2000global},
locally linear embedding~\cite{roweis2000nonlinear},
Laplacian eigenmaps~\cite{belkin2003laplacian},
diffusion maps~\cite{coifman2006diffusion}
and t-SNE~\cite{maaten2008visualizing}.
Diffusion maps is used in this paper.

The input for the dimensionality reduction is a set of
feature vectors $\bm{F}_i(x,y) (i=1,\ldots,N)$,
which are created from the local image patches around a pixel $(x,y)$
of frames $i=1,\ldots,N$. In the experiments,
the size of a patch is $11\times11$ pixels and the dimension of
the feature vector is $11^2$. 
The speckle pattern of a image patch is embedded into $\bm{\Psi}_i(x,y)$ 
by the dimensionality reduction.
The embedding is applied for each pixel independently
and $N$ vectors of dimension $l$ are obtained as the outputs.
The dimension $l$ is $5$ in the experiments.

\subsection{Making the embedded vectors spatially consistent}
\label{sec:spatial}

The embedded vector $\bm{\Psi}_i(x,y)$ has ambiguity in its sign
because the speckle has no information about the direction of the movement.
Since the embedded vectors are calculated independently for each pixel
by the method described in \secref{embedding},
the parameters of adjacent pixels can be inconsistent each other.
The second step of the proposed method is to make them spatially consistent.

A linear transformation is introduced to modify the embedded vectors
as follows:
\begin{align}
  \bm{\psi}_i(x,y) = \bm{M}(x,y) \bm{\Psi}_i(x,y)
  \label{equ:psi}
\end{align}
where $\bm{M}(x,y)$ is a $3 \times l$ matrix 
and $\bm{\Psi}_i(x,y)$ is a $l \times 1$ column vector for the pixel $(x,y)$.
Since the local plane parameters of neighbor pixels should be similar,
the following constraint is satisfied to make $\bm{\psi}_i(x,y)$ spatially consistent
\begin{align}
  \min_{\bm{M}} & \sum_i \sum_{(x_1,y_1)} \sum_{(x_2,y_2)} E_s(i,x_1,y_1,x_2,y_2) \\
  E_s(i,x_1,y_1,x_2,y_2) = 
  \parallel & [x_2 - x_1, y_2 - y_1, 1] \bm{M}(x_1,y_1) \bm{\Psi}_i(x_1,y_1) - \notag \\
  & [0, 0, 1] \bm{M}(x_2,y_2) \bm{\Psi}_i(x_2,y_2) \parallel^2 \label{equ:E_s}
\end{align}
where $(x_1,y_1)$ is a point in the whole image and 
$(x_2,y_2)$ is a point in the neighborhood window of $(x_1,y_1)$.
This constraint means that the 3D local patches in \figref{speckle_example_modeling}(c) have
the same height at the overlapping pixels in the image.
The constraint for a pair of $(x_1,y_1)$ and $(x_2,y_2)$ becomes
\begin{align}
  (\bm{\Psi}_i(x_1,y_1) [x_2 - x_1, y_2 - y_1, 1]) \circ \bm{M}(x_1,y_1) & \notag   \\
  - (\bm{\Psi}_i(x_2,y_2) [0, 0, 1]) \circ \bm{M}(x_2,y_2)               & = \bm{0}
  \label{equ:constraint}
\end{align}
where $\circ$ is element-wise product.
Let $\bm{m}$ the column vector by stacking $\bm{M}(x,y)$ for all pixels after vectorization.
Since the constraints are expressed by $\bm{A} \bm{m} = \bm{0}$, where
$\bm{A}$ is the coefficient matrix calculated from \equref{constraint},
$\bm{m}$ is given as the eigenvector associated with the smallest eigenvalue of $\bm{A}^T \bm{A}$.
Once $\bm{M}(x,y)$ is calculated, the modified embedded vectors $\bm{\psi}_i(x,y)$
calculated by \equref{psi} become spatially consistent.
Note that the ambiguity of the movement direction in total remains
due to the ambiguity of the eigenvector even after making it spatially consistent.

One of the problem in calculating the eigenvector is that
the temporal distribution of $\bm{\Psi}_i(x,y)$ between pixels is different.
If $\bm{\Psi}_i(x,y)$ of a pixel $(x,y)$ is almost zero,
the magnitude of $\bm{M}(x,y)$ is nearly equal to one
and the parameters for other pixels become almost zero,
because $\bm{\psi}_i(x,y)$ does not affect the error
even if the magnitude of $\bm{M}(x,y)$ is dominant in the eigenvector.
Therefore, the standard deviation of $\bm{\Psi}_i(x,y)$ is normalized
before calculating the coefficient matrix $\bm{A}$ along the time axis
as follows:
\begin{align}
\Psi'_{ik} = \Psi_{ik} / (\sigma_k + \epsilon), \quad
\sigma^2_k = \frac{1}{N} \sum_i (\Psi_{ik} - \overline{\Psi_k})^2
\end{align}
where $\Psi_{ik}$ is the $k$-th component of $\bm{\Psi}_i$,
$\overline{\Psi_k}$ is the mean value of $\Psi_{ik}$ and 
$\epsilon$ is a small number to avoid division by zero.

Since the cost of calculating the eigenvector in high resolution is large,
$\bm{M}(x,y)$ is calculated after subsampling pixels.
In the experiments, the original image is $512 \times 512$ pixels and
subsampled to $64 \times 64$ pixels.

\subsection{Optimizing surface parameters}

The calculation of $\bm{M}(x,y)$ in \secref{spatial} does not
consider the distance between the features calculated by \equref{distance}.
In the third step, $\bm{M}(x,y)$ is optimized so that
the embedded vector preserves the distance of the features.

By assuming the movement between adjacent two frames is sufficiently small,
the constraint so that the distance of the embedded vectors preserves
the distance of the features is added to the error function to be minimized
as follows:
\begin{align}
  \min_{\bm{M}} \sum_i \left[ \sum_{(x,y)} E_t(i,x,y) + 
  \lambda \sum_{(x_1,y_1)} \sum_{(x_2,y_2)} E_s(i,x_1,y_1,x_2,y_2) \right] \label{equ:optim}\\
  E_t(i,x,y) = (\parallel \bm{\psi}_i(x,y) - \bm{\psi}_{i+1}(x,y) \parallel -
  \parallel \bm{F}_i(x,y) - \bm{F}_{i+1}(x,y) \parallel)^2
\end{align}
where $E_s$ is the spatial constraint defined by \equref{E_s} and
$\lambda$ is its weight.
Once the transformation matrix $\bm{M}(x,y)$ is optimized, 
the plane parameter $\bm{\psi}_i(x,y)$ is obtained, and
The displacement for each pixel is given as the third component 
of $\bm{\psi}_i(x,y)$.

The initial guess $\bm{\psi}'_i(x,y)$ for nonlinear minimization of \equref{optim}
is given by $\bm{\psi}_i(x,y)$ calculated in \secref{spatial}
after normalizing the magnitude for each pixel as follows:
\begin{align}
  \bm{\psi}'_i(x,y) &= \frac{m_F}{m_\psi + \epsilon} \bm{\psi}_i(x,y) \\
  m_\psi &= \frac{1}{N} \sum_i \parallel \bm{\psi}_i(x,y) - \bm{\psi}_{i+1}(x,y) \parallel)^2 \\
  m_F &= \frac{1}{N} \sum_i \parallel \bm{F}_i(x,y) - \bm{F}_{i+1}(x,y) \parallel)^2
\end{align}

Since $\bm{M}(x,y)$ calculated in \secref{spatial} is subsampled,
$\bm{M}(x,y)$ is interpolated before optimizing parameters
to obtain the displacement for the image of the original resolution.
It is done by interpolating the displacement of the subsampled images
calculated by using the initial guess $\bm{\psi}'_i(x,y)$.
Let $d(x,y)$ the linearly interpolated displacement at the pixel $(x,y)$
of the original resolution.
$\bm{M}(x,y)$ at the interpolated pixels should satisfy the following equation
for the neighboring pixels $(\tilde{x}, \tilde{y})$:
\begin{align}
  [\tilde{x} - x, \tilde{y} - y, 1] \bm{M}(x,y) \bm{\Psi}_i(x,y) = d(\tilde{x}, \tilde{y})
\end{align}
Since this linear equation of $\bm{M}(x,y)$ is obtained for all pixels
in the local patch, $\bm{M}(x,y)$ is calculated as the least-square solution.
Once $\bm{M}(x,y)$ is obtained for all pixels,
the optimization is applied with the images of the original resolution.



\section{Experiments}

In the experiments, we used a high-speed monochrome camera that captures images
at more than 1000 frames/second and a laser light source as the incident light
to produce the speckle pattern. The resolution of the camera is $512 \times 512$ pixels.
The light source and camera have about 40 degrees field of view.

\subsection{Evaluating the calculated displacement with respect to the real offset}

\begin{figure}[t]
  \begin{center}
    \begin{tabular}{ccc}
      \includegraphics[width=0.3\textwidth]{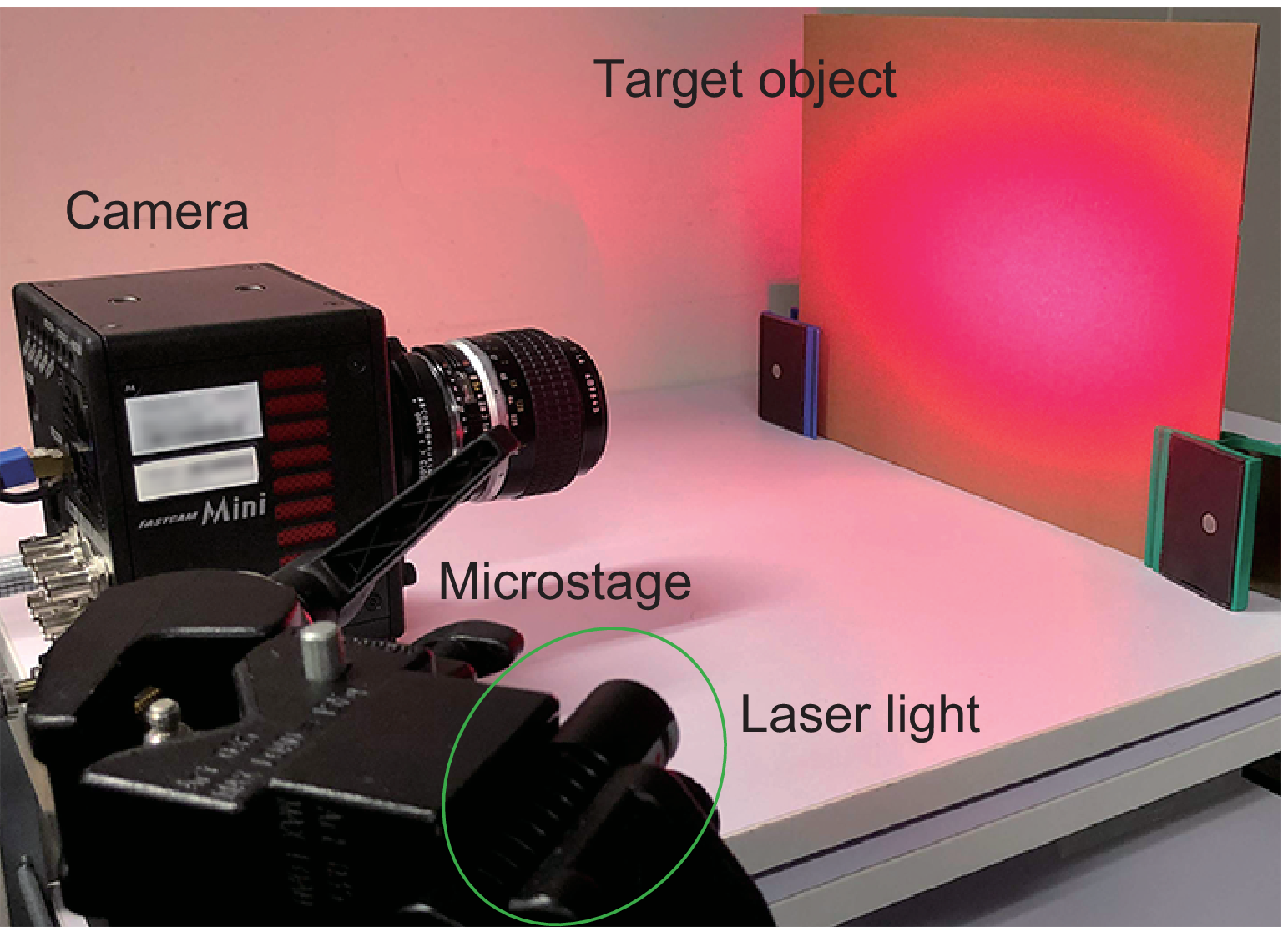} &
      {
        \scriptsize
        \begin{tabular}[b]{ccc}
          \includegraphics[width=0.093\textwidth]{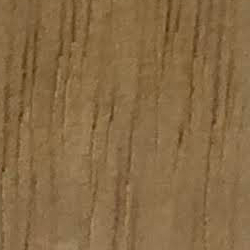} &
          \includegraphics[width=0.093\textwidth]{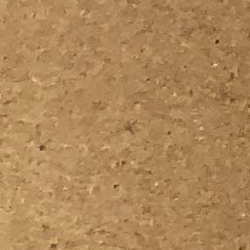} &
          \includegraphics[width=0.093\textwidth]{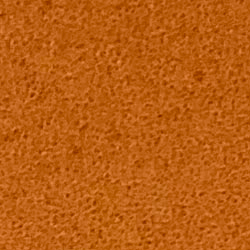} \\
          \includegraphics[width=0.093\textwidth]{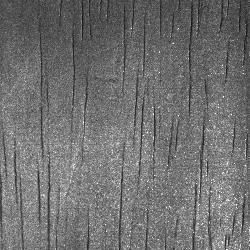} &
          \includegraphics[width=0.093\textwidth]{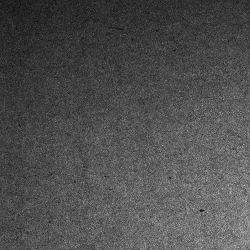} &
          \includegraphics[width=0.093\textwidth]{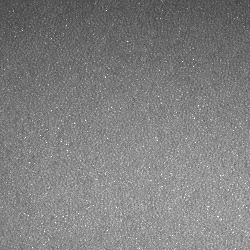} \\
          Wood & Cardboard & Sponge
        \end{tabular}
      } &
      \includegraphics[width=0.33\textwidth]{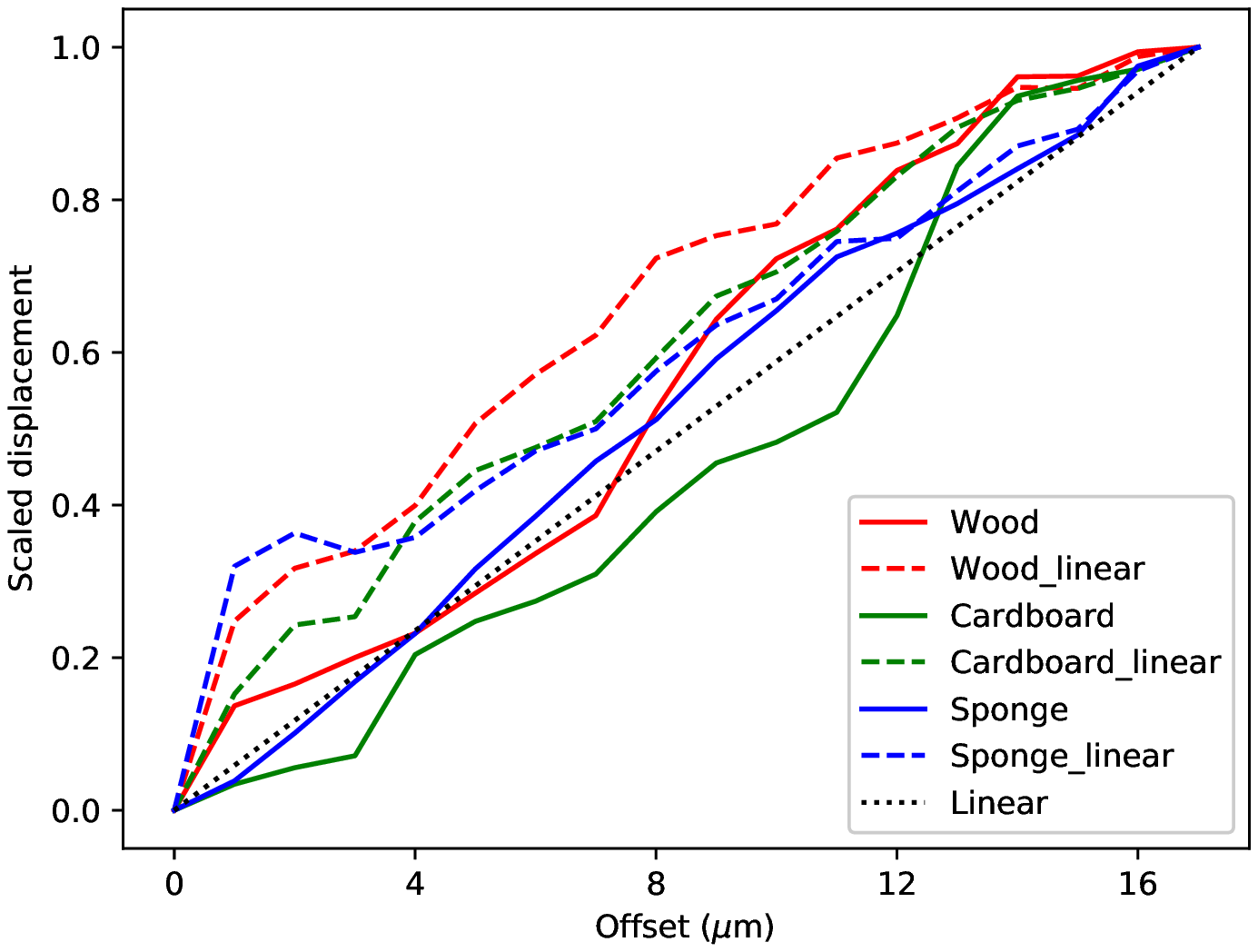} \\
      (a) & (b) & (c) \\
    \end{tabular}
  \end{center}
  \caption{ (a) A planar object is illuminated by a laser light source and
  observed by a camera mounted on a microstage, which moves the camera along the
  axis. (b) Three materials (wood, cardboard, sponge) are tested. The images in
  the upper row are illuminated by room light, and the ones in the lower row are
  illuminated by laser light. (c) The relationship between the calculated
  displacements and the real offsets are shown. The displacements are scaled
  from 0.0 to 1.0.}
  \label{fig:microstage_exp}
\end{figure}


The first experiment is to evaluate the displacement calculated by the proposed method
with the images by changing the position with known offsets.
\figref{microstage_exp}(a) shows the experimental setup.
A planar object is illuminated by a laser projector and is observed by a camera.
The camera is mounted on a microstage and the position is moved
along the camera axis. A set of 100 images are obtained in total
by capturing five images at each offset during changing the offset 
from 0 to \um{18} every \um{1}.
The proposed method described in Sections \ref{sec:embedding} and \ref{sec:spatial}
is applied to the set of 100 images since the order of images 
cannot be used for the optimization for this dataset.
In this experiment, we evaluate if the displacement calculated
by the proposed method has linear relationship compared to the offset
given by the microstage.
Since speckle can be observed if an object has rough surface,
three types of materials (wood, cardboard and sponge) are tested in this experiment
to evaluate the relationship between the calculated displacement and the real offset.




\figref{microstage_exp}(c) shows the results of the calculated displacements.
Since the absolute values of displacements can not be obtained by the proposed method,
the values are scaled from 0.0 to 1.0, which correspond 
from 0 to \um{18}
of the real offset. The displacements are averaged at each offset.
The results are compared with the linear approach calculating Euclidean distance 
between the patches of zero offset and the others.
The root-mean-square errors of the displacements from the linear relationship 
are 0.080, 0.076 and 0.038 for the three materials, wood, cardboard and sponge,
in the scaled displacement.
The values correspond to \um{1.36}, \um{1.28} and \um{0.64} in the real scale.
The root-mean-square errors of the linear approach are 0.171, 0.108 and 0.118 in the scaled displacement, 
and \um{2.91}, \um{1.83} and \um{2.01} in the real scale, respectively.
These results show the proposed method is applicable for various materials
to discriminate the minute displacements
and the relationship with respect to the real movement is close to linear than
by using linear distance.

\subsection{Comparison with accelerometers}

\begin{figure}[t]
  \begin{center}
    \begin{tabular}{cc}
      \includegraphics[height=0.16\textheight]{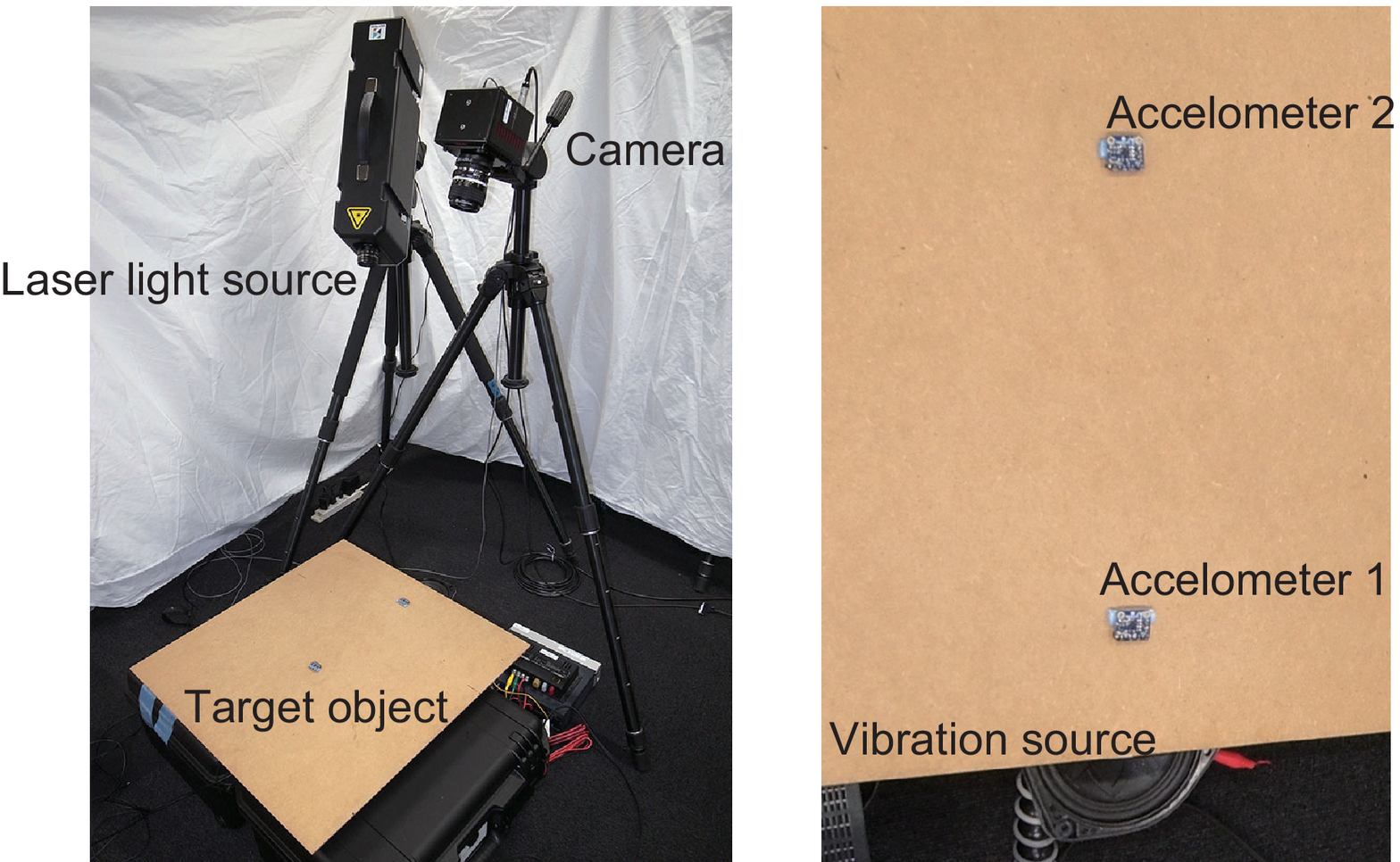} &
      \includegraphics[height=0.16\textheight]{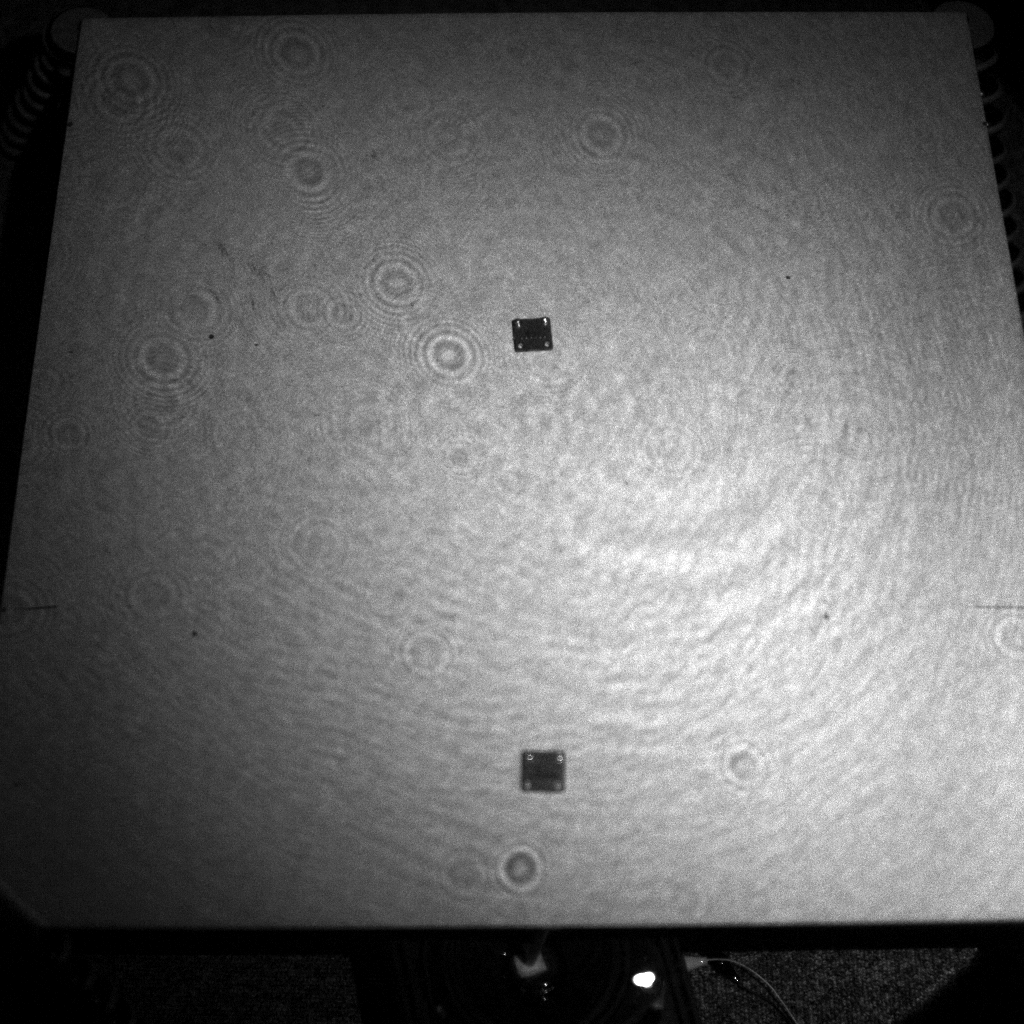}
    \end{tabular}
  \end{center}
  \caption{The experimental setup consists of a camera and laser light
    source. A speaker is used as a source to vibrate the target
    object.  Two accelerometers are placed on the object to measure the
    vibration. The right image is one of input images illuminated by
    the laser light.}
  \label{fig:experimental_setup}
\end{figure}

\begin{figure*}[t]
  \begin{center}
   \tabcolsep = 0.1mm
   \scriptsize
    \begin{tabular}{ccccc}
     \includegraphics[width=0.19\textwidth]{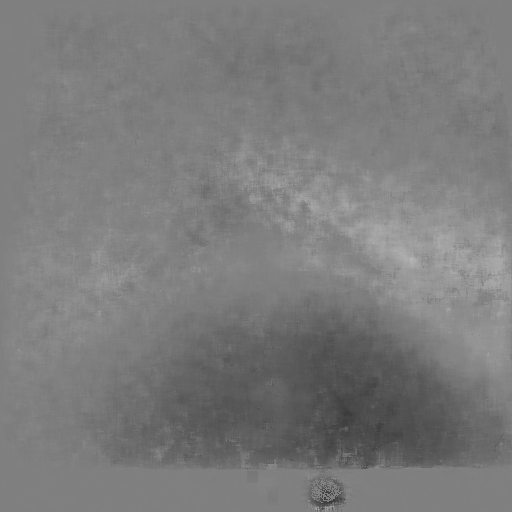} &
     \includegraphics[width=0.19\textwidth]{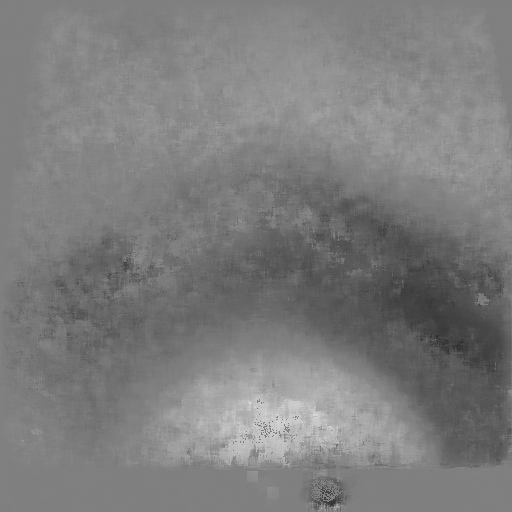} &
     \includegraphics[width=0.19\textwidth]{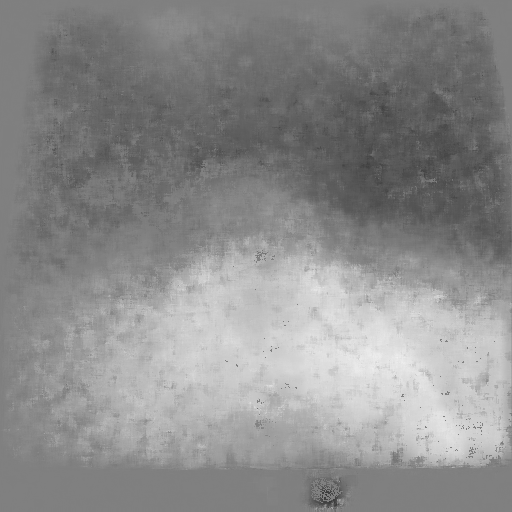} &
     \includegraphics[width=0.19\textwidth]{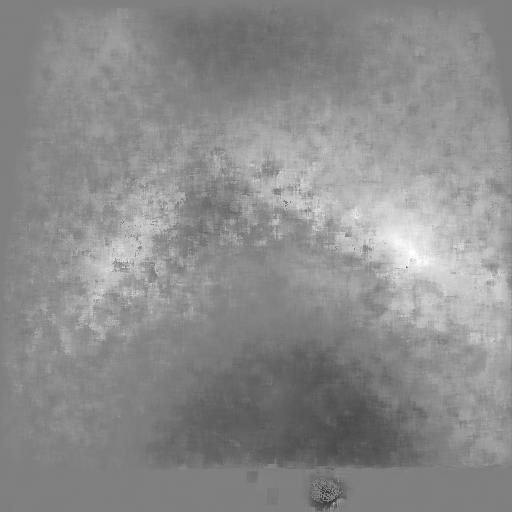} &
     \includegraphics[width=0.19\textwidth]{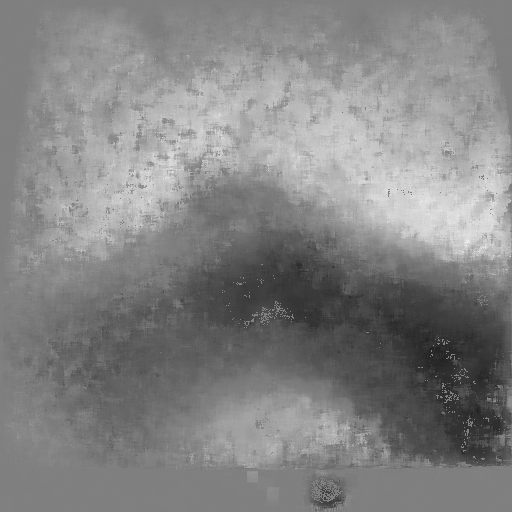} \\
     Frame 52 & Frame 57 & Frame 62 & Frame 67 & Frame 72
    \end{tabular}
  \end{center}
  \caption{The calculated displacements are shown for five frames in the sequence.
  The displacements are indicated by the brightness in these images.
  }
  \label{fig:pulse1_imgs}
\end{figure*}

\begin{figure}[t]
  \begin{center}
    \includegraphics[width=1.0\textwidth]{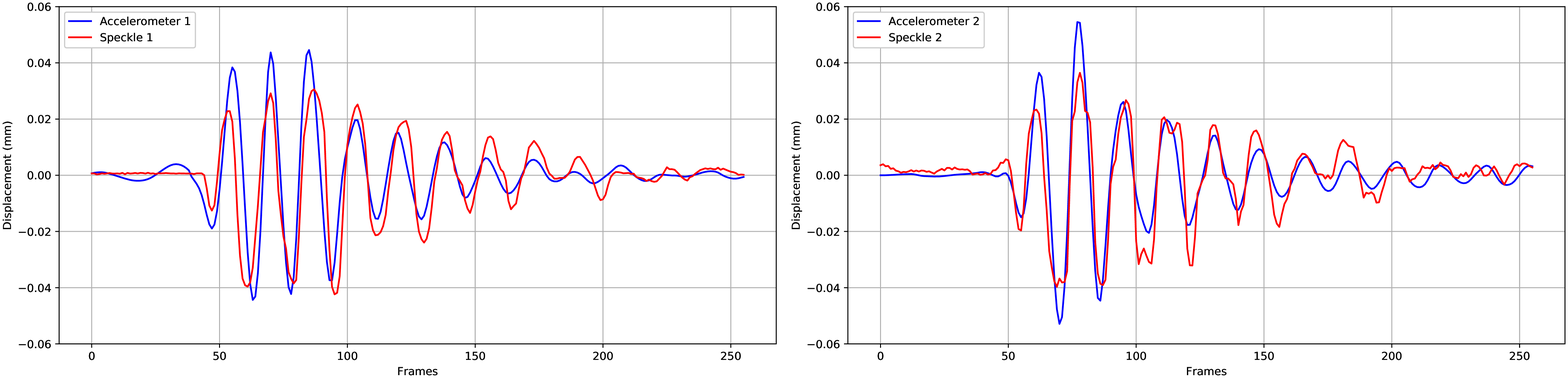}
  \end{center}
  \caption{The measurements by two accelerometers are compared to the
    results by the proposed method. 
    }
  \label{fig:compare_accel_wood_pulse1}
\end{figure}

\begin{figure}[t]
  \begin{tabular}{cc}
    \begin{minipage}{0.35\textwidth}
      \begin{center}
        \includegraphics[width=1.0\textwidth]{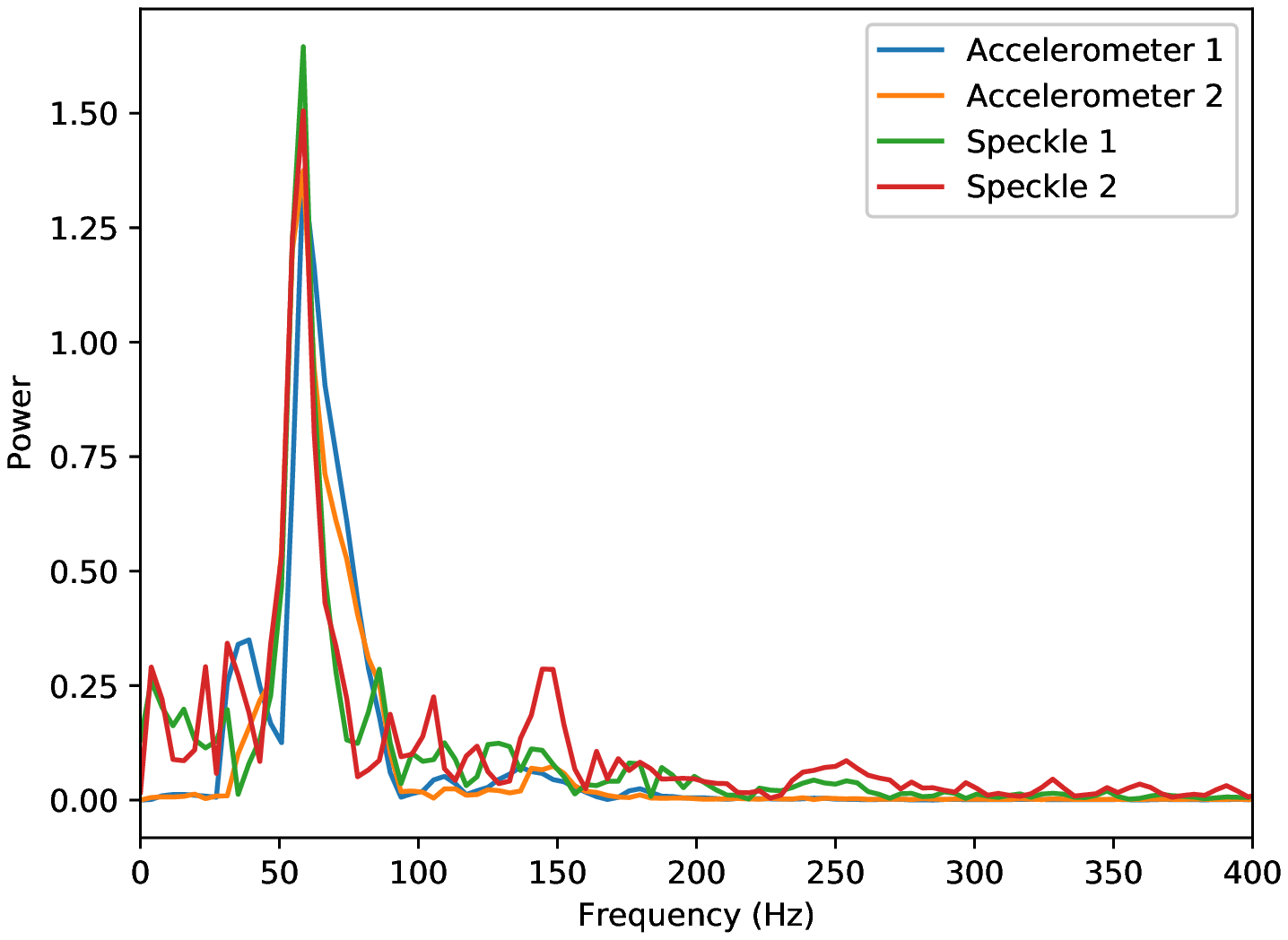}
      \end{center}
      \caption{The power spectrum}
      \label{fig:compare_accel_wood_pulse1_freq}
    \end{minipage}
    &
    \begin{minipage}{0.65\textwidth}
      \begin{center}
        \scriptsize
        \begin{tabular}{cc}
          \includegraphics[width=0.5\textwidth]{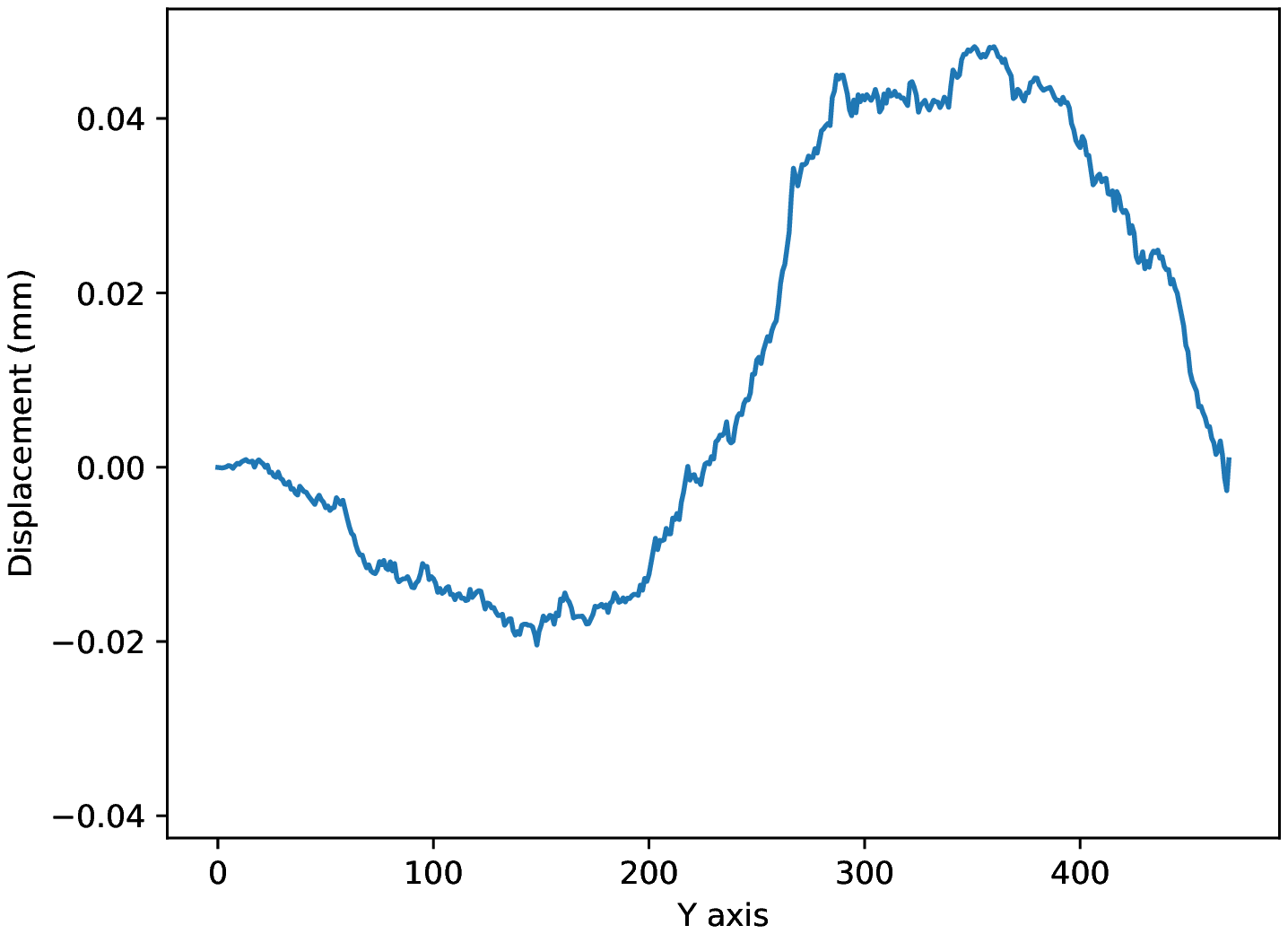} &
          \includegraphics[width=0.5\textwidth]{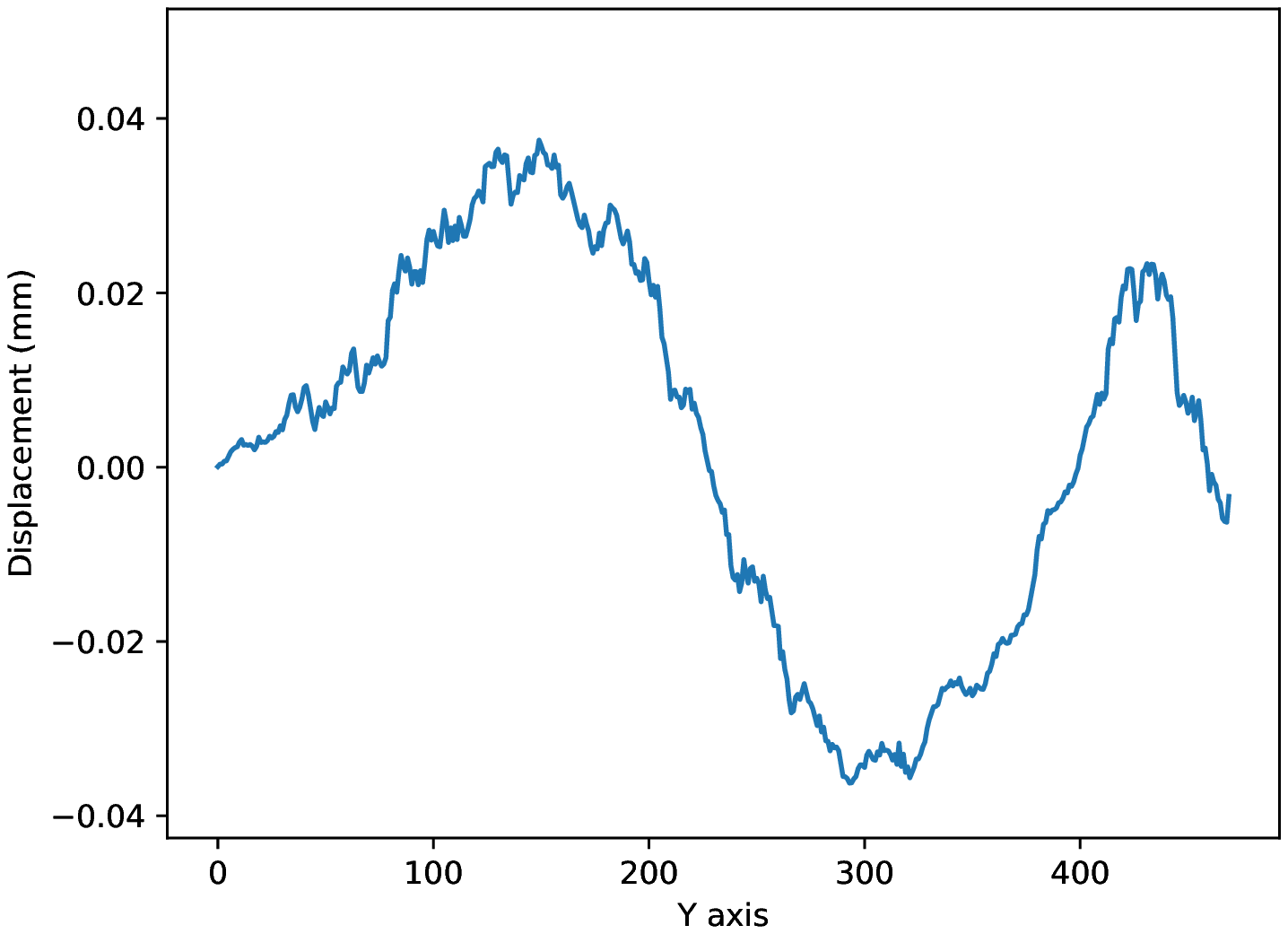} \\
          Frame 62 & Frame 72
        \end{tabular}
      \end{center}
      \caption{The vertical slices of the displacements}
      \label{fig:pulse1_x}
    \end{minipage}
  \end{tabular}
\end{figure}

Second, the proposed method is tested for a case of dynamic movement.
The result is compared to the measurements by the accelerometers.
\figref{experimental_setup} shows the setup in the experiments. The
distance from the camera to the target object (wood panel) is about 1m, and the
field of view is about 0.6m square at the target's position. The
object is in contact with a speaker as the vibration source, which
generates a 60Hz sine wave. The direction of the vibration is nearly
parallel to the viewing direction and the movement is out-of-plane. 
The image sequence of 256 frames is captured by a high speed camera at 1000 frames/second. 
Two accelerometers are placed on the object to measure the
vibration and compare with the results by the proposed method.


\figref{pulse1_imgs} shows the results of calculated displacement
of five frames in the image sequence.
The area of the same brightness indicates the same displacement.
The vibration propagated from the bottom of the image at the beginning,
and became a standing wave in the latter part.
The measurements by the proposed method is
compared to the values calculated by integrating the acceleration
twice. \figref{compare_accel_wood_pulse1} shows the comparison with two
measurements. Since the proposed method cannot calculate the absolute
values of displacement, it is scaled to fit the mean magnitude of the
measurements by the accelerometers. The two
accelerometers are placed at the different distance from the vibration
source. The accelerometer 2 is delayed about the half of the cycle from
the accelerometer 1. The results by the proposed method is synchronized
with the measurements by the both accelerometers.  The normalized cross
correlation of the signals between the proposed method and the
accelerometers are 0.768 and 0.863. The high correlation indicates the
proposed method can extract the phase of the minute vibration from the
speckle pattern. \figref{compare_accel_wood_pulse1_freq} shows
the power spectrum of the vibration measured by the accelerometers
and the proposed method. A strong peak at the same frequency
is observed in all the results.
\figref{pulse1_x} shows the vertical slices of the displacements
along the center line of the images of the two frames 
in \figref{pulse1_imgs}. 
Since the spatial consistency of the displacements calculated 
for each pixel independently in \secref{embedding} 
is obtained by the method described in \secref{spatial},
the wave propagated in the image can be observed by the displacements.

\subsection{Visualizing various movements}

\begin{figure*}[t]
  \begin{center}
  \tabcolsep = 0.1mm
  \scriptsize
    \begin{tabular}{ccccc}
      \includegraphics[width=0.19\textwidth]{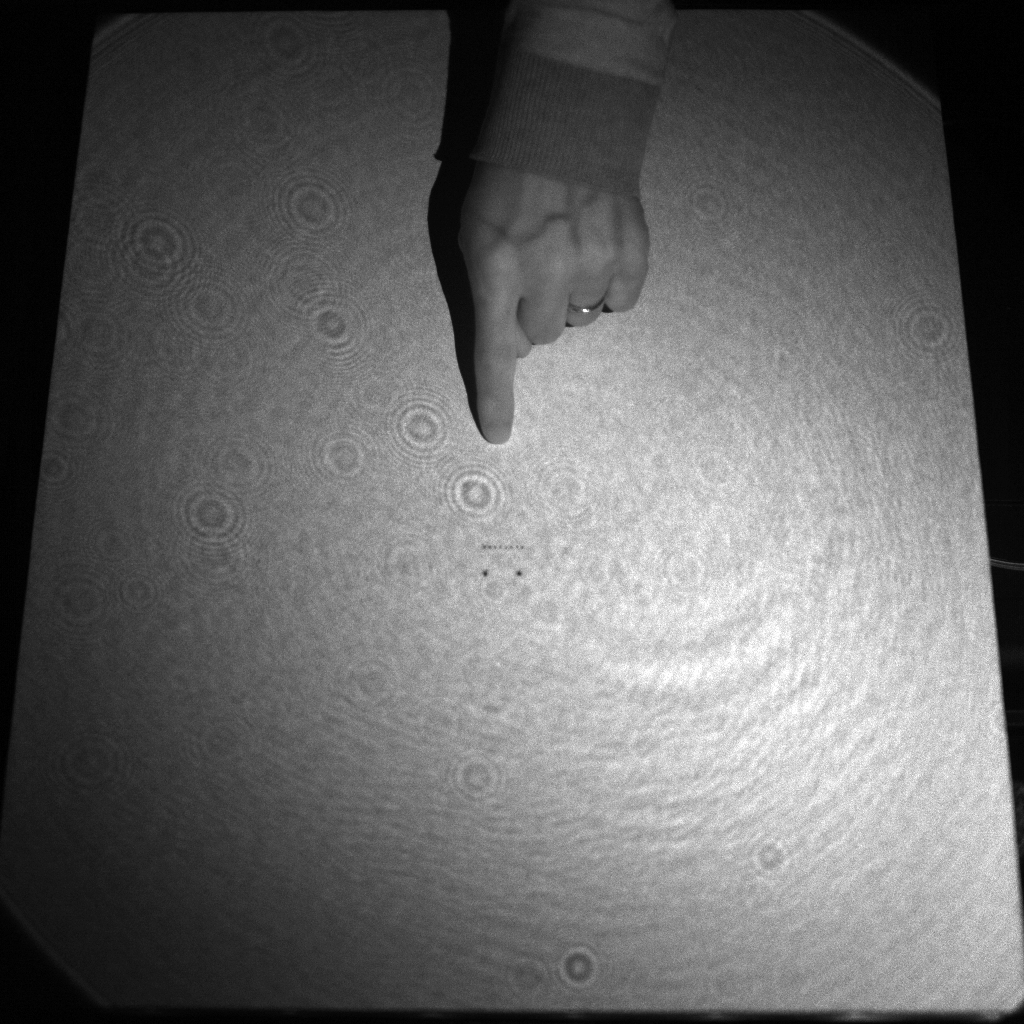} &
      \includegraphics[width=0.19\textwidth]{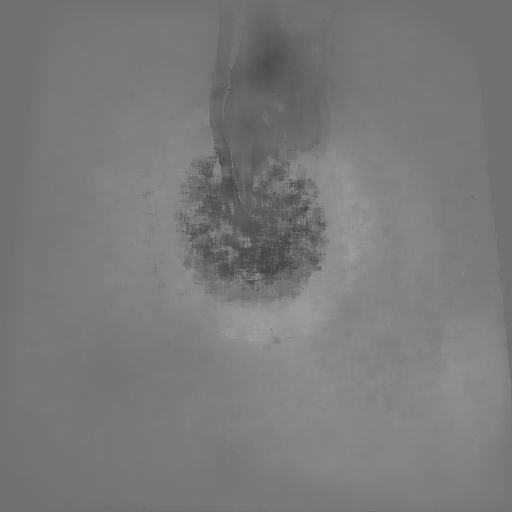} &
      \includegraphics[width=0.19\textwidth]{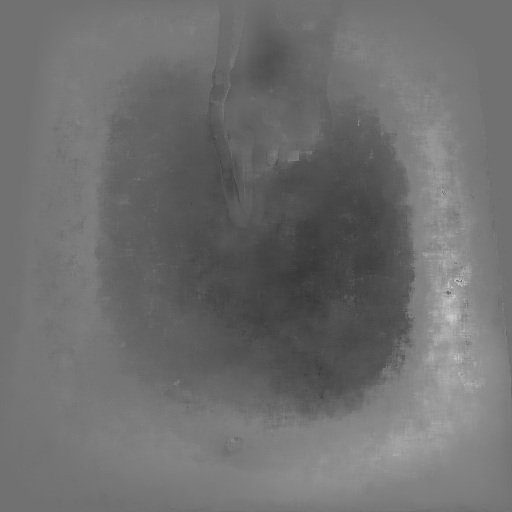} &
      \includegraphics[width=0.19\textwidth]{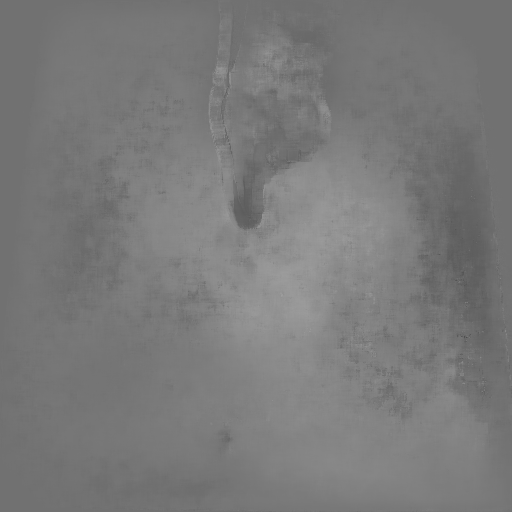} &
      \includegraphics[width=0.19\textwidth]{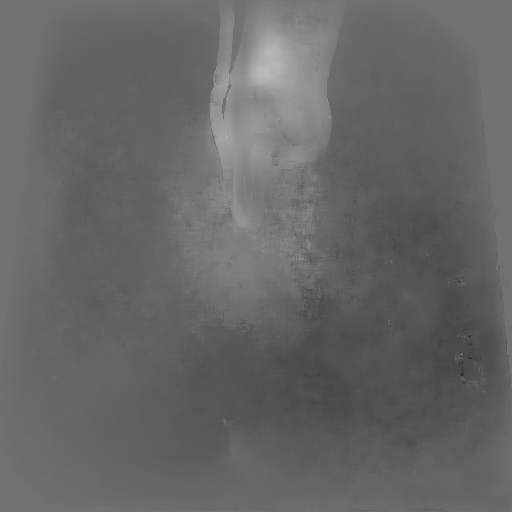} \\
      Input & 6ms & 16ms & 129ms & 190ms
    \end{tabular}
  \end{center}
  \caption{The movement is produced by touching the canvas by a
    finger in this sequence. The movement starts from the touching
    point, and propagates as a circular wave. 
    }
  \label{fig:touch_sequence}
\end{figure*}

\ifx
\begin{figure*}[t]
  \begin{center}
  \scriptsize
    \begin{tabular}{ccc}
     \includegraphics[width=0.3\textwidth]{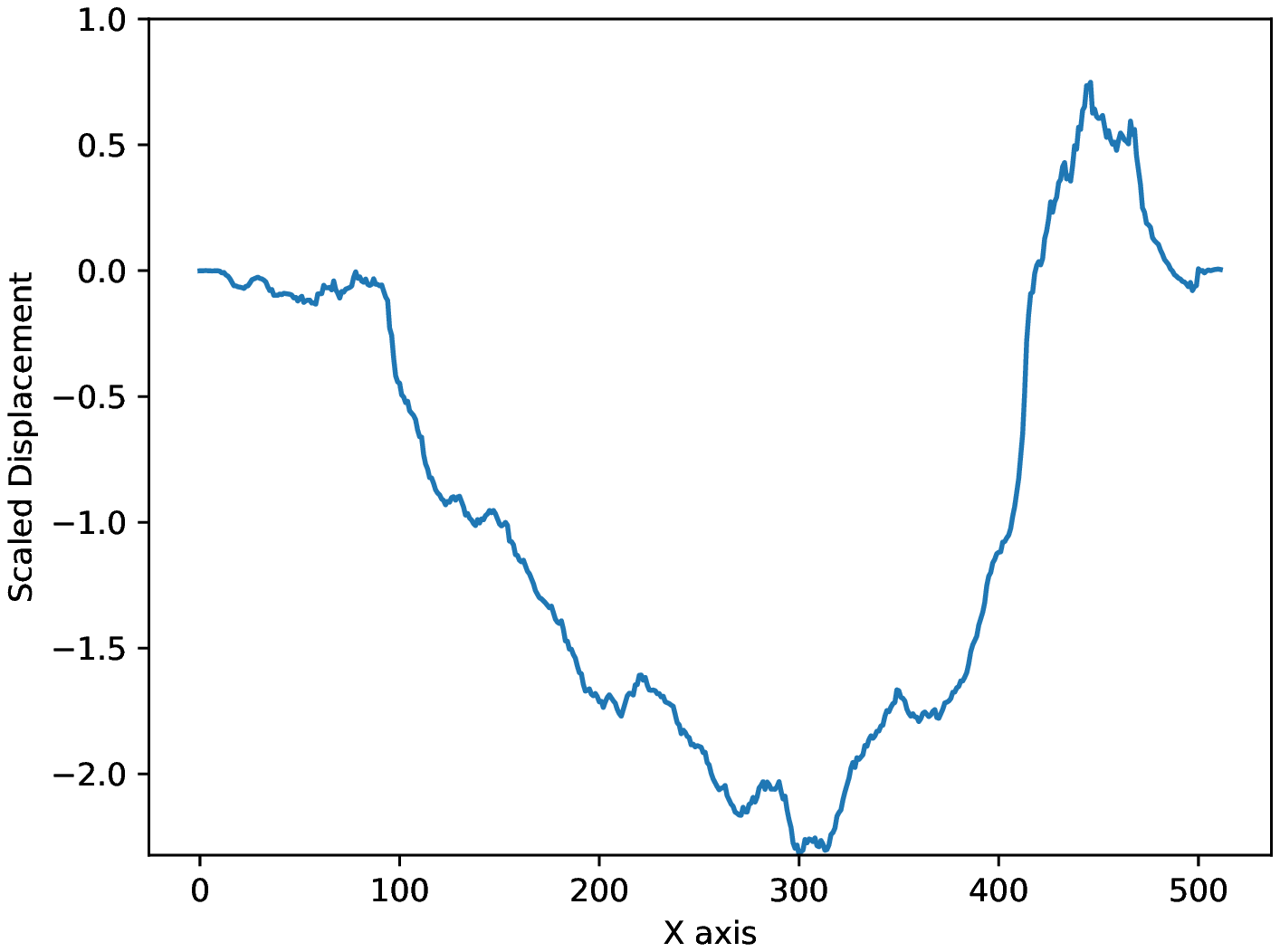} &
     \includegraphics[width=0.3\textwidth]{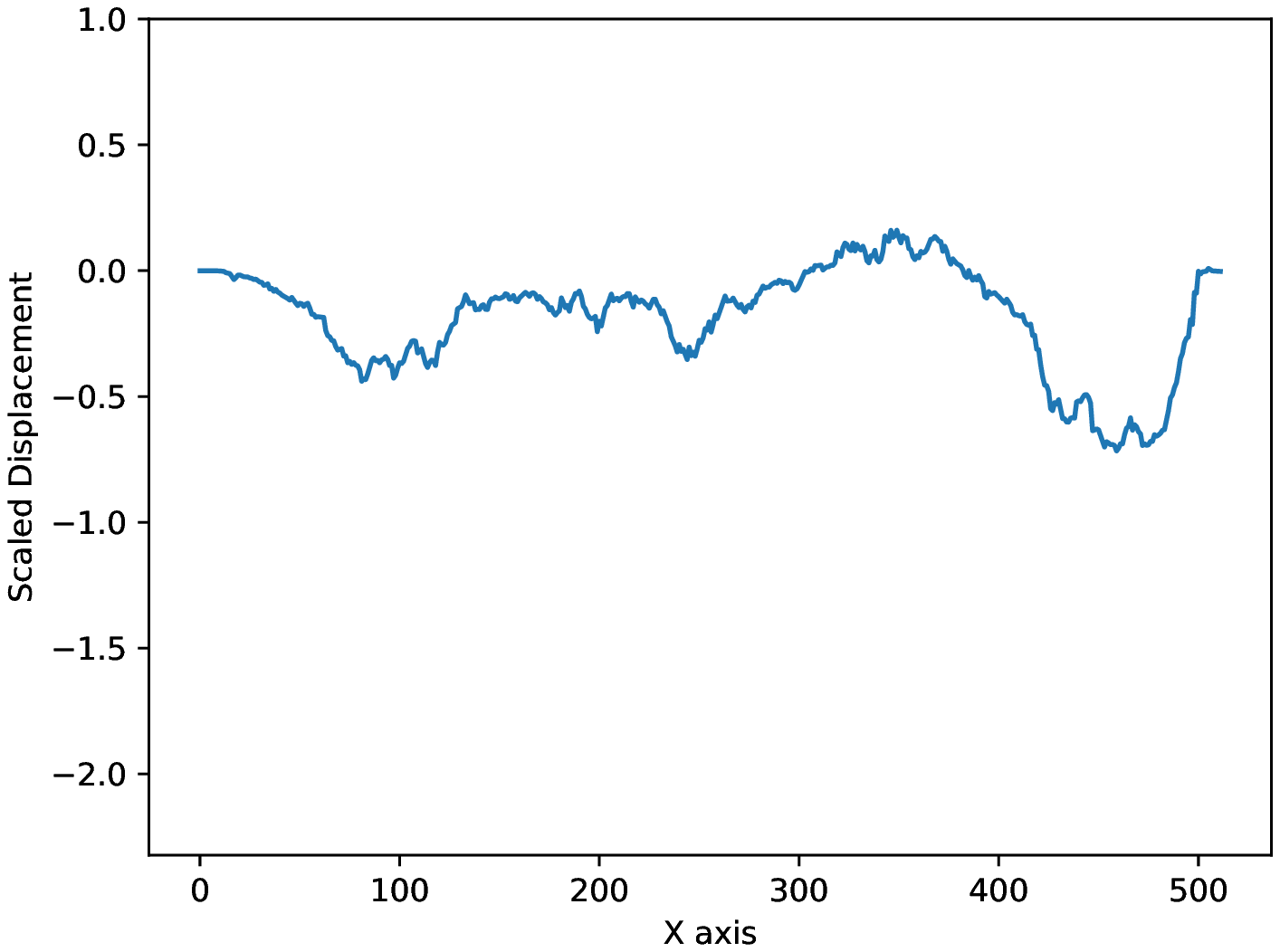} &
     \includegraphics[width=0.3\textwidth]{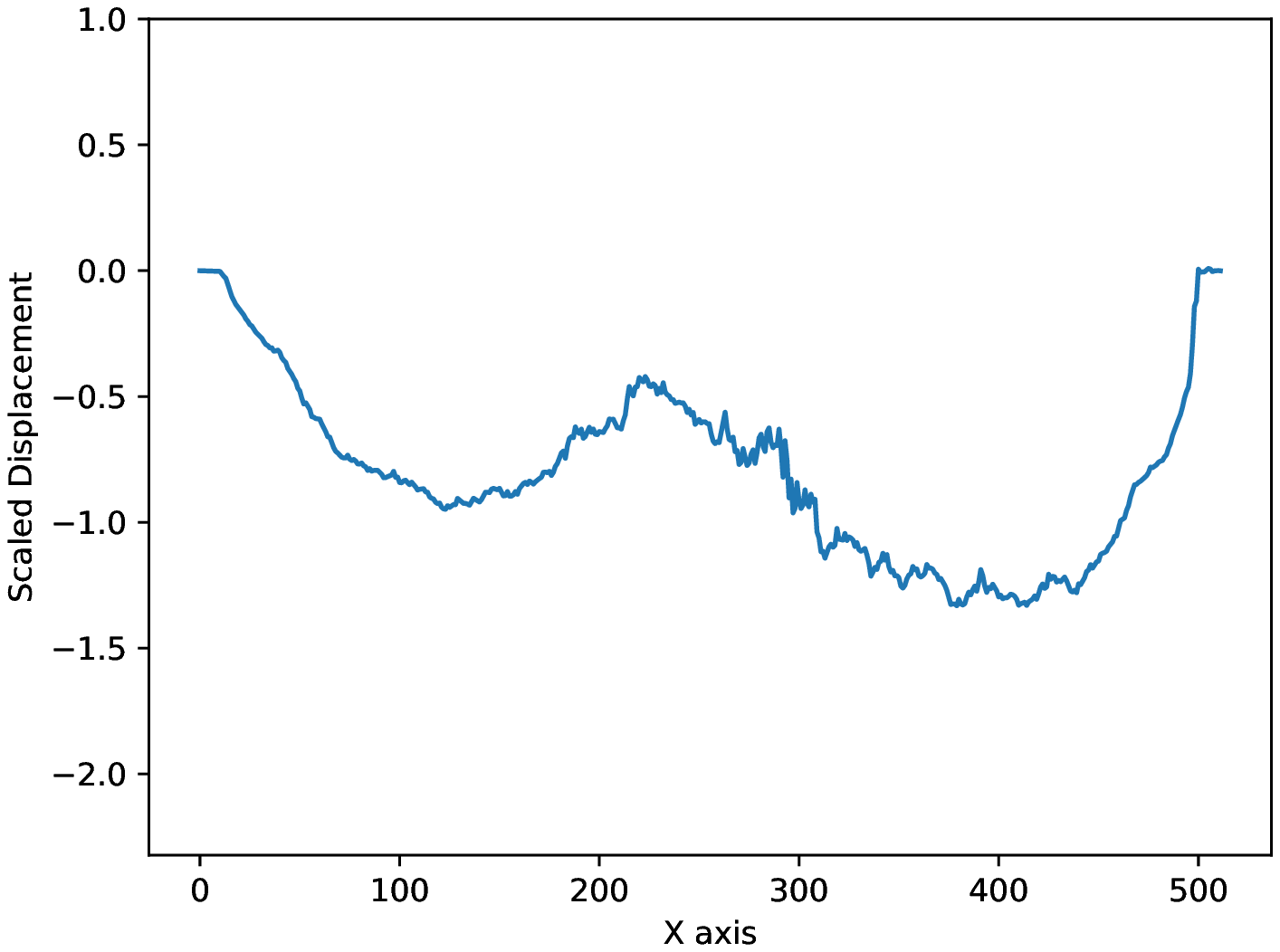} \\
     16ms & 129ms & 190ms
    \end{tabular}
  \end{center}
  \caption{The horizontal slice of the displacements of the three frames
  in \figref{touch_sequence} are shown.}
  \label{fig:touch_graph}
\end{figure*}
\fi

\ifx
\begin{wrapfigure}{r}[10pt]{0.5\textwidth}
  \centering
      \includegraphics[height=0.16\textheight]{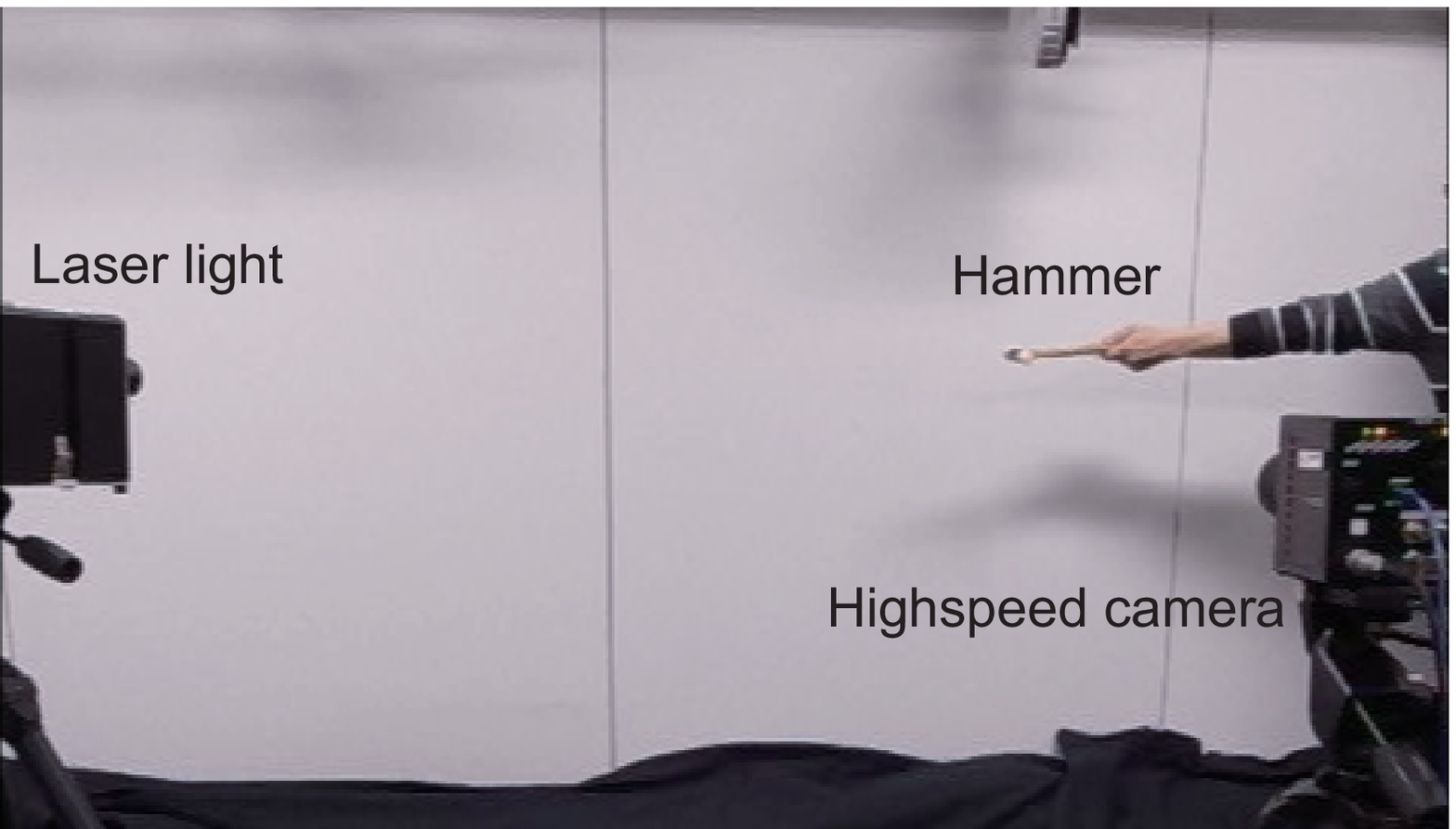}
  \caption{The experiment of hitting a painted metal wall by a hammer is observed by a high speed camera.
  The right image is one of the input images illuminated by infrared laser light.}
  \label{fig:wall_exp}
\end{wrapfigure}
\fi
Third, the proposed method is applied to various movements of different materials.
In \figref{touch_sequence}, a finger touches the canvas cloth and the movement occurs 
from the touching points. The movement propagates as a circular wave, 
which is reflected at the edge of the canvas.
The left image in \figref{touch_sequence} shows one of the image
sequences of the input images and the rest is the resulting displacement images.
The images are captured at 1000 frames / second in this experiment.
The canvas is pushed down by the finger at first
and the position returns back after releasing the touch.
The canvas cloth vibrates repeatedly during the touch.
\ifx
\begin{figure}[t]
  \begin{center}
    \scriptsize
    \begin{tabular}{cc}
      \includegraphics[height=0.16\textheight]{wall_exp.eps} &
      \includegraphics[height=0.16\textheight]{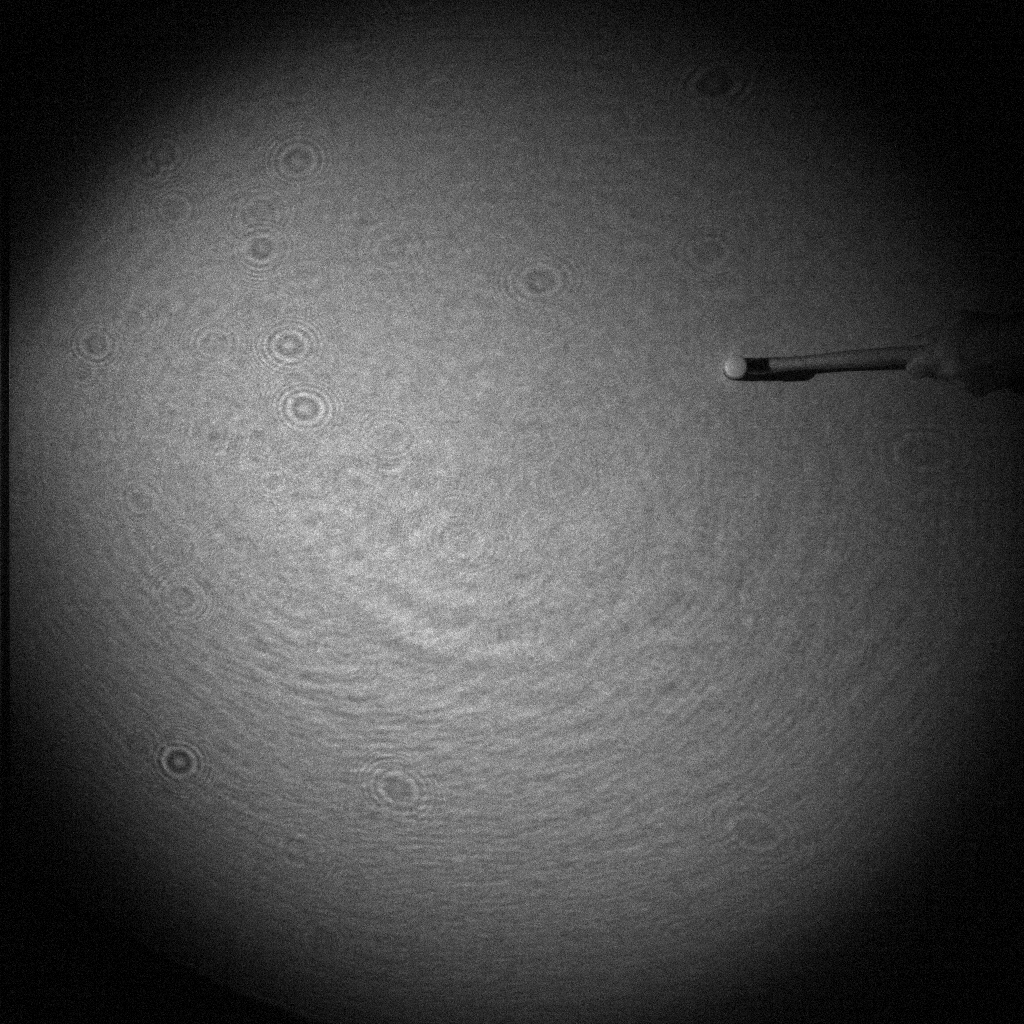} \\
      Experimental setup & Input image
    \end{tabular}
  \end{center}
  \caption{The experiment of hitting a painted metal wall by a hammer is observed by a high speed camera.
  The right image is one of the input images illuminated by infrared laser light.}
  \label{fig:wall_exp}
\end{figure}
\fi

\begin{figure}[t]
  \begin{center}
    \scriptsize
    \begin{tabular}{cc}
      \includegraphics[height=0.13\textheight]{wall_exp.eps} &
      \includegraphics[height=0.13\textheight]{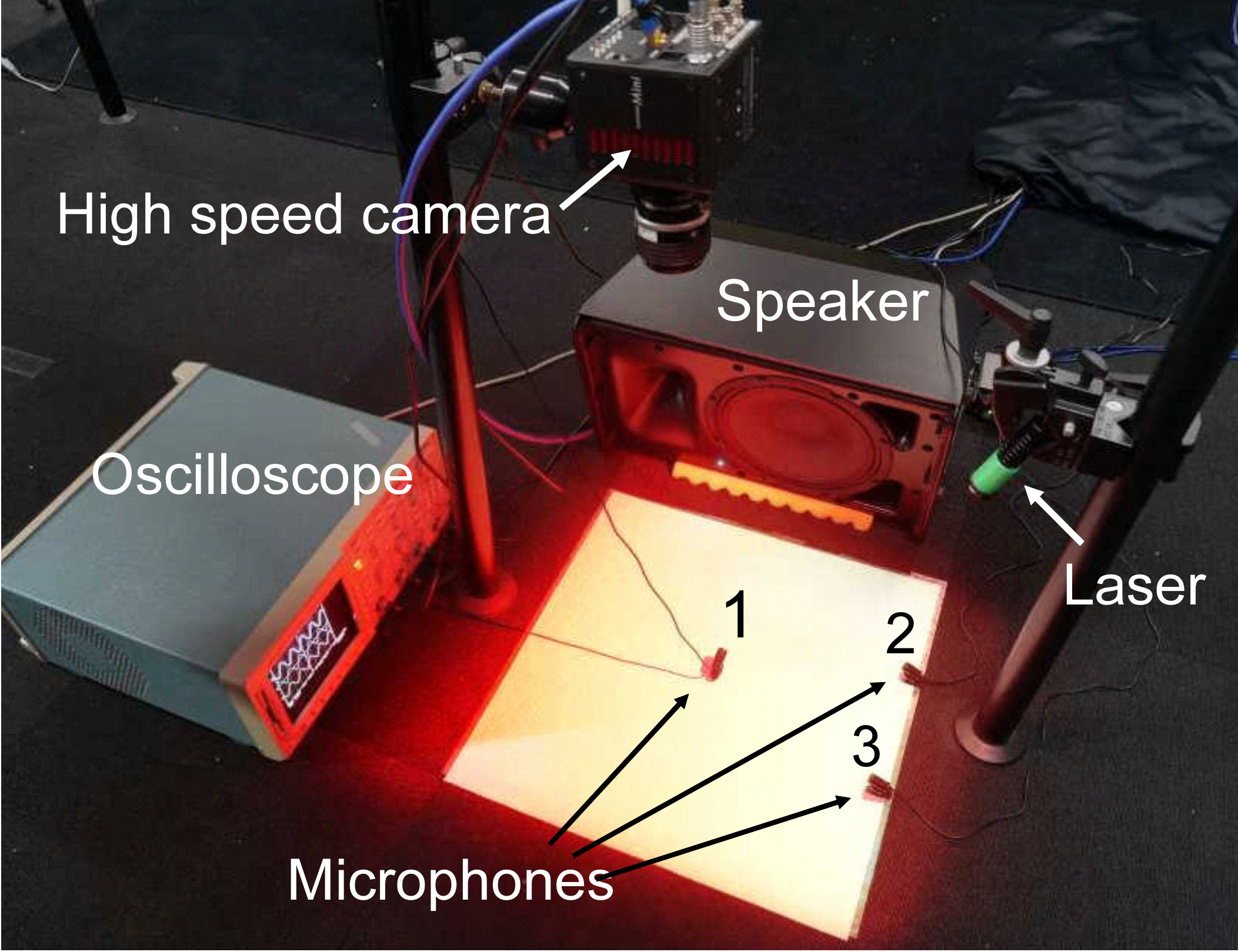} \\
      (a) & (b)
    \end{tabular}
  \end{center}
  \caption{Experimental setups: (a) The experiment of hitting a painted metal wall by a hammer.
  (b) The experiment of visualizing sound wave.}
  \label{fig:wall_exp}
\end{figure}

\renewcommand{\topfraction}{1.0}
\renewcommand{\bottomfraction}{1.0}
\renewcommand{\dbltopfraction}{1.0}
\renewcommand{\textfraction}{0.01}
\renewcommand{\floatpagefraction}{1.0}
\renewcommand{\dblfloatpagefraction}{1.0}
\setcounter{topnumber}{5}
\setcounter{bottomnumber}{5}
\setcounter{totalnumber}{10}

\begin{figure*}[t]
  \begin{center}
    \begin{tabular}{ccccc}
      \includegraphics[width=0.19\textwidth]{wall_input0135.png} &
      \includegraphics[width=0.19\textwidth]{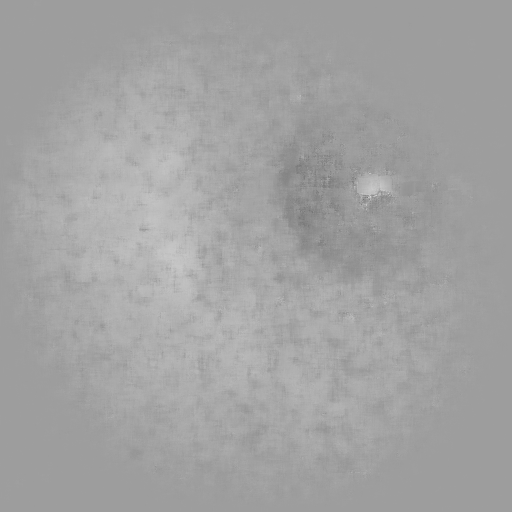} &
      \includegraphics[width=0.19\textwidth]{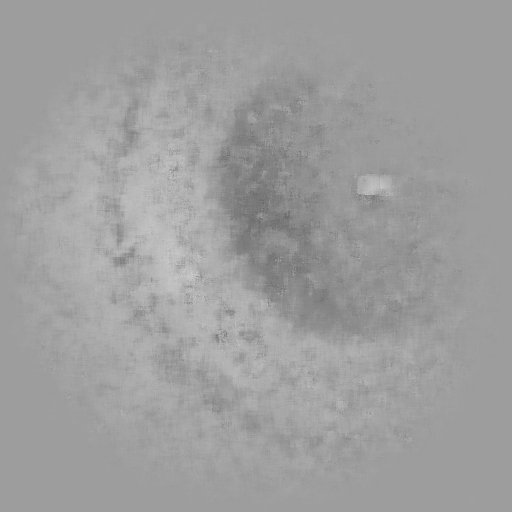} &
      \includegraphics[width=0.19\textwidth]{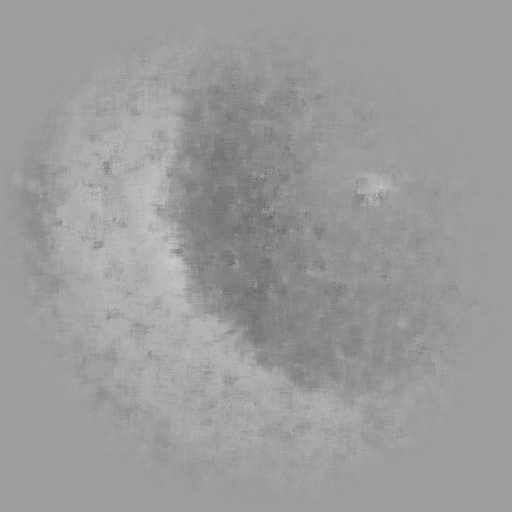} &
      \includegraphics[width=0.19\textwidth]{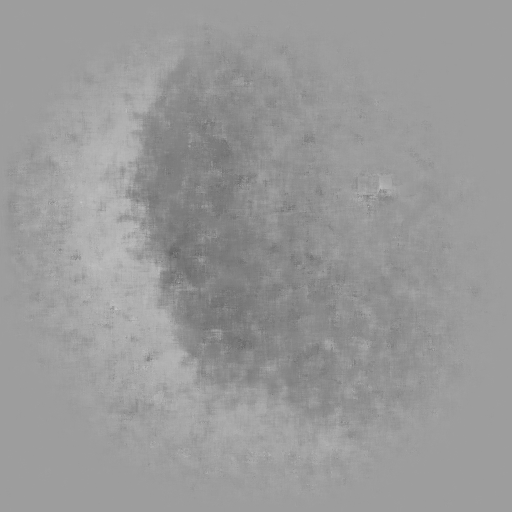} \\
     Input & 0.63ms & 1.41ms & 2.19ms & 2.97ms
    \end{tabular}
  \end{center}
  \caption{The displacements generated by hammering are calculated.
  }
  \label{fig:wall_sequence}
\end{figure*}

Next, a hammering experiment of a wall made of metal panel is observed by the proposed method.
The wall is painted and the speckle can be observed on the surface.
\figref{wall_exp}(a) shows the situation and one of the input images.
In this experiment, the camera equips the band-pass filter to capture
the infrared laser light and discard the illumination by room light.
The image sequence is captured at 6400 frames/second in this experiment.
\figref{wall_sequence} shows the displacements for four frames in the sequence.
The wave started from the hitting point and propagated the metal wall.
Since the size of wall is known and the speed of the wave is
about 17.5 pixel/frame, the speed of wave that propagates
in the metal wall is about \SI{250}{m/s}.

\ifx
\begin{wrapfigure}{r}[10pt]{0.37\textwidth}
  \centering
    \includegraphics[width=0.35\textwidth]{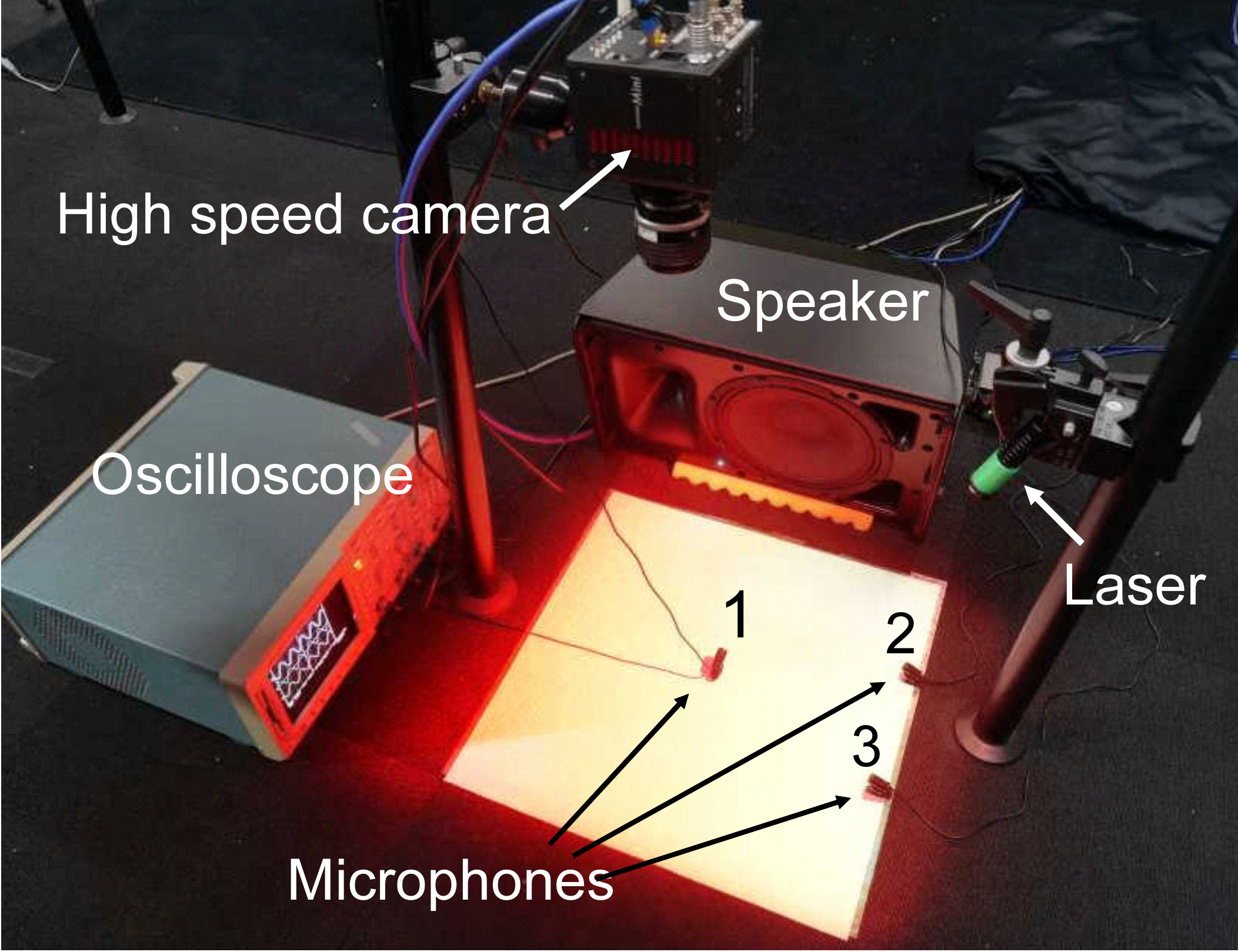}
 \caption{Setup.}
 \label{fig:floortissue_scene}
\end{wrapfigure}
\fi

\ifx
\begin{figure}[bt]
  \begin{center}
\ifx
    \includegraphics[width=0.35\textwidth]{figs/sound_cap_scene.pdf}\\
(a) Setup\\
    \begin{tabular}{cccccc}
      \includegraphics[width=0.16\textwidth]{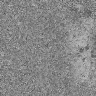} &
      \includegraphics[width=0.16\textwidth]{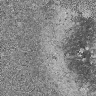} &
      \includegraphics[width=0.16\textwidth]{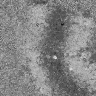} &
      \includegraphics[width=0.16\textwidth]{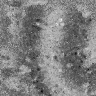} &
      \includegraphics[width=0.16\textwidth]{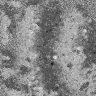} &
      \includegraphics[width=0.16\textwidth]{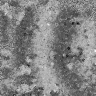} \\
      \includegraphics[width=0.16\textwidth]{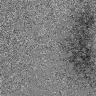} &
      \includegraphics[width=0.16\textwidth]{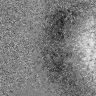} &
      \includegraphics[width=0.16\textwidth]{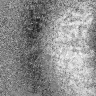} &
      \includegraphics[width=0.16\textwidth]{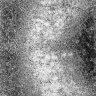} &
      \includegraphics[width=0.16\textwidth]{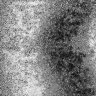} &
      \includegraphics[width=0.16\textwidth]{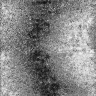} \\
      \includegraphics[width=0.16\textwidth]{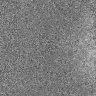} &
      \includegraphics[width=0.16\textwidth]{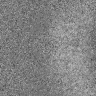} &
      \includegraphics[width=0.16\textwidth]{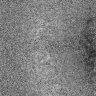} &
      \includegraphics[width=0.16\textwidth]{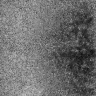} &
      \includegraphics[width=0.16\textwidth]{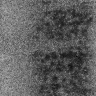} &
      \includegraphics[width=0.16\textwidth]{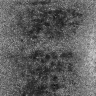} \\
      \includegraphics[width=0.16\textwidth]{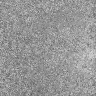} &
      \includegraphics[width=0.16\textwidth]{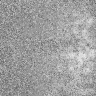} &
      \includegraphics[width=0.16\textwidth]{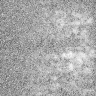} &
      \includegraphics[width=0.16\textwidth]{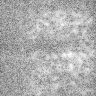} &
      \includegraphics[width=0.16\textwidth]{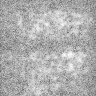} &
      \includegraphics[width=0.16\textwidth]{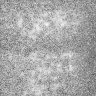} \\
    \end{tabular}
Same frame number interval\\
\fi
    \begin{tabular}{ccccccc}
	\begin{minipage}{0.02\hsize}
  	\rotatebox{90}{200Hz}
	\end{minipage}&
      \includegraphics[width=0.15\textwidth]{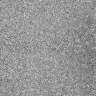} &
      \includegraphics[width=0.15\textwidth]{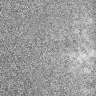} &
      \includegraphics[width=0.15\textwidth]{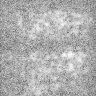} &
      \includegraphics[width=0.15\textwidth]{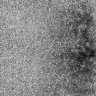} &
      \includegraphics[width=0.15\textwidth]{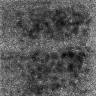} &
      \includegraphics[width=0.15\textwidth]{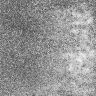} \\
      & frame \#46 & \#56 & \#66 & \#76 & \#86 & \#96 \\
	\begin{minipage}{0.02\hsize}
  	\rotatebox{90}{500Hz}
	\end{minipage}&
      \includegraphics[width=0.15\textwidth]{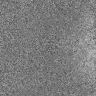} &
      \includegraphics[width=0.15\textwidth]{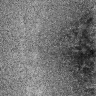} &
      \includegraphics[width=0.15\textwidth]{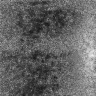} &
      \includegraphics[width=0.15\textwidth]{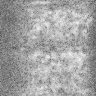} &
      \includegraphics[width=0.15\textwidth]{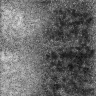} &
      \includegraphics[width=0.15\textwidth]{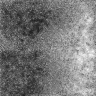} \\
      & frame \#49 & \#57 & \#65 & \#73 & \#81 & \#89 \\
	\begin{minipage}{0.02\hsize}
  	\rotatebox{90}{1000Hz}
	\end{minipage}&
      \includegraphics[width=0.15\textwidth]{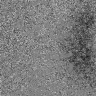} &
      \includegraphics[width=0.15\textwidth]{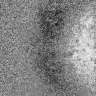} &
      \includegraphics[width=0.15\textwidth]{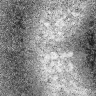} &
      \includegraphics[width=0.15\textwidth]{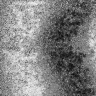} &
      \includegraphics[width=0.15\textwidth]{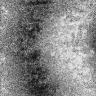} &
      \includegraphics[width=0.15\textwidth]{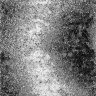} \\
      & frame \#51 & \#55 & \#59 & \#63 & \#67 & \#74 \\
	\begin{minipage}{0.02\hsize}
  	\rotatebox{90}{2000Hz}
	\end{minipage}&
      \includegraphics[width=0.15\textwidth]{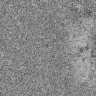} &
      \includegraphics[width=0.15\textwidth]{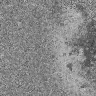} &
      \includegraphics[width=0.15\textwidth]{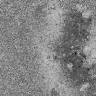} &
      \includegraphics[width=0.15\textwidth]{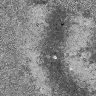} &
      \includegraphics[width=0.15\textwidth]{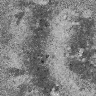} &
      \includegraphics[width=0.15\textwidth]{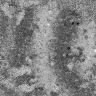} \\
      & frame \#35 & \#37 & \#39 & \#41 & \#43 & \#45 \\
    \end{tabular}
(a) Estimated displacement maps for different frequencies of sounds.\\
    \includegraphics[width=0.26\textwidth]{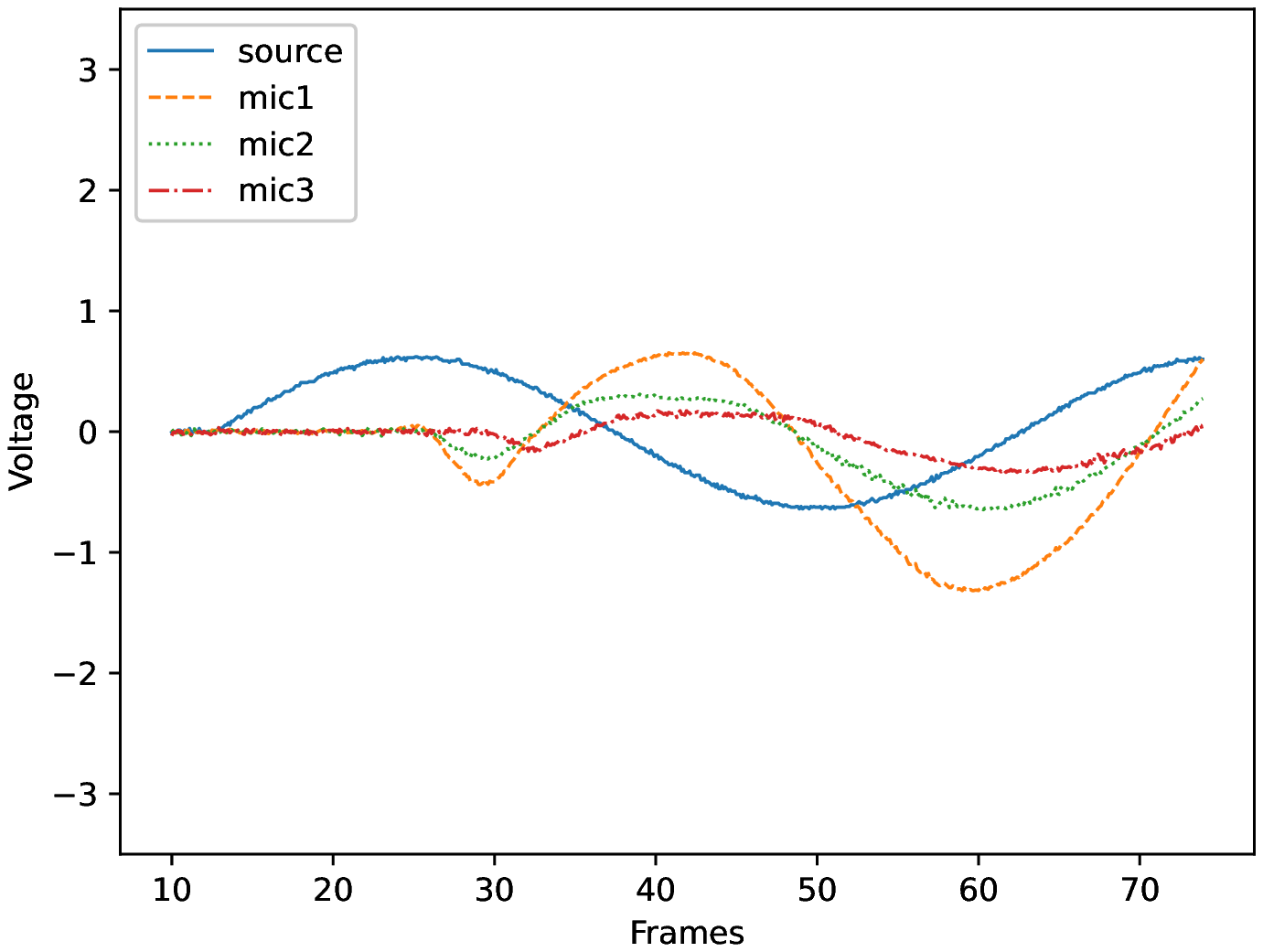}
    \includegraphics[width=0.26\textwidth]{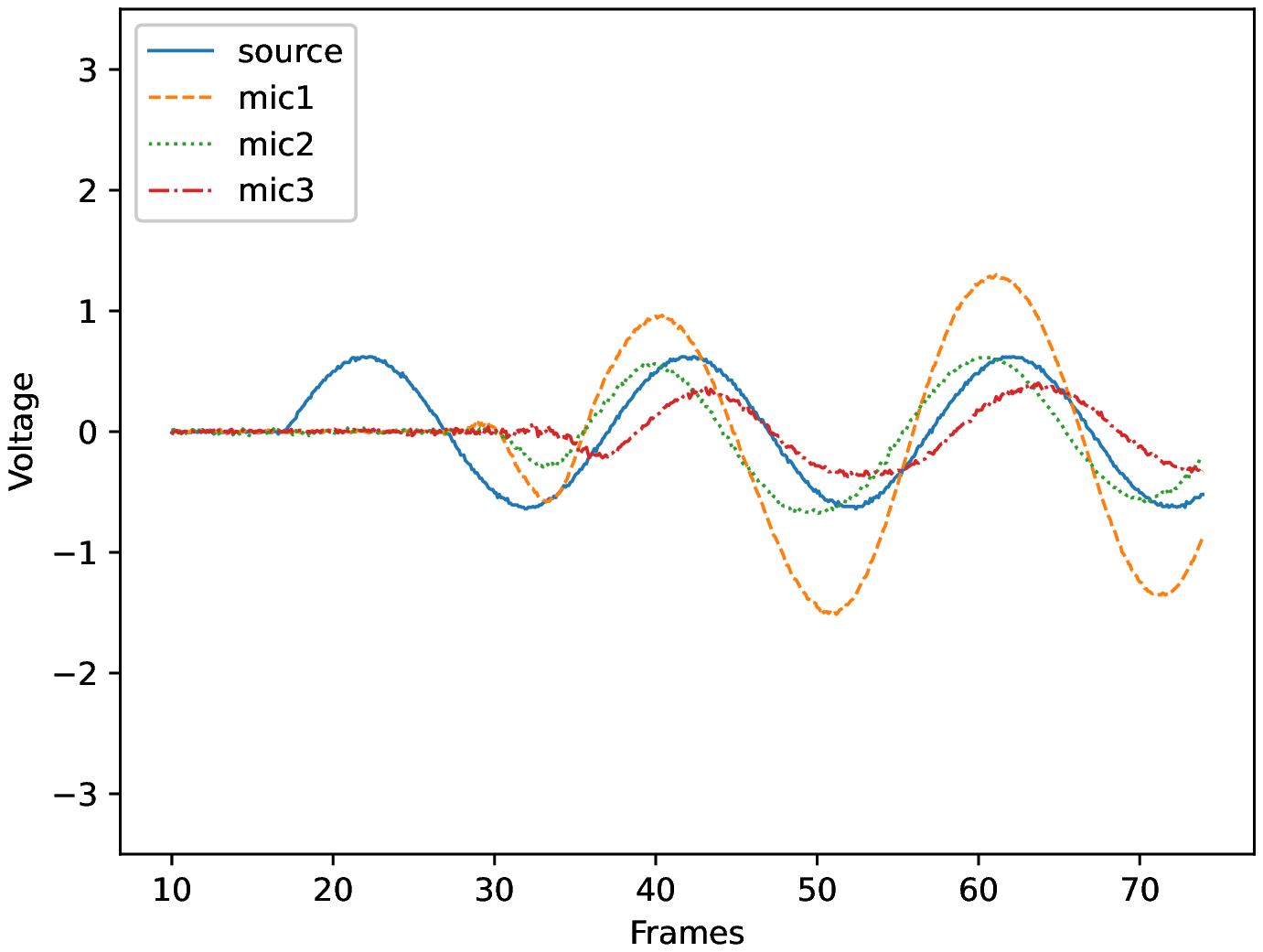}
    \includegraphics[width=0.26\textwidth]{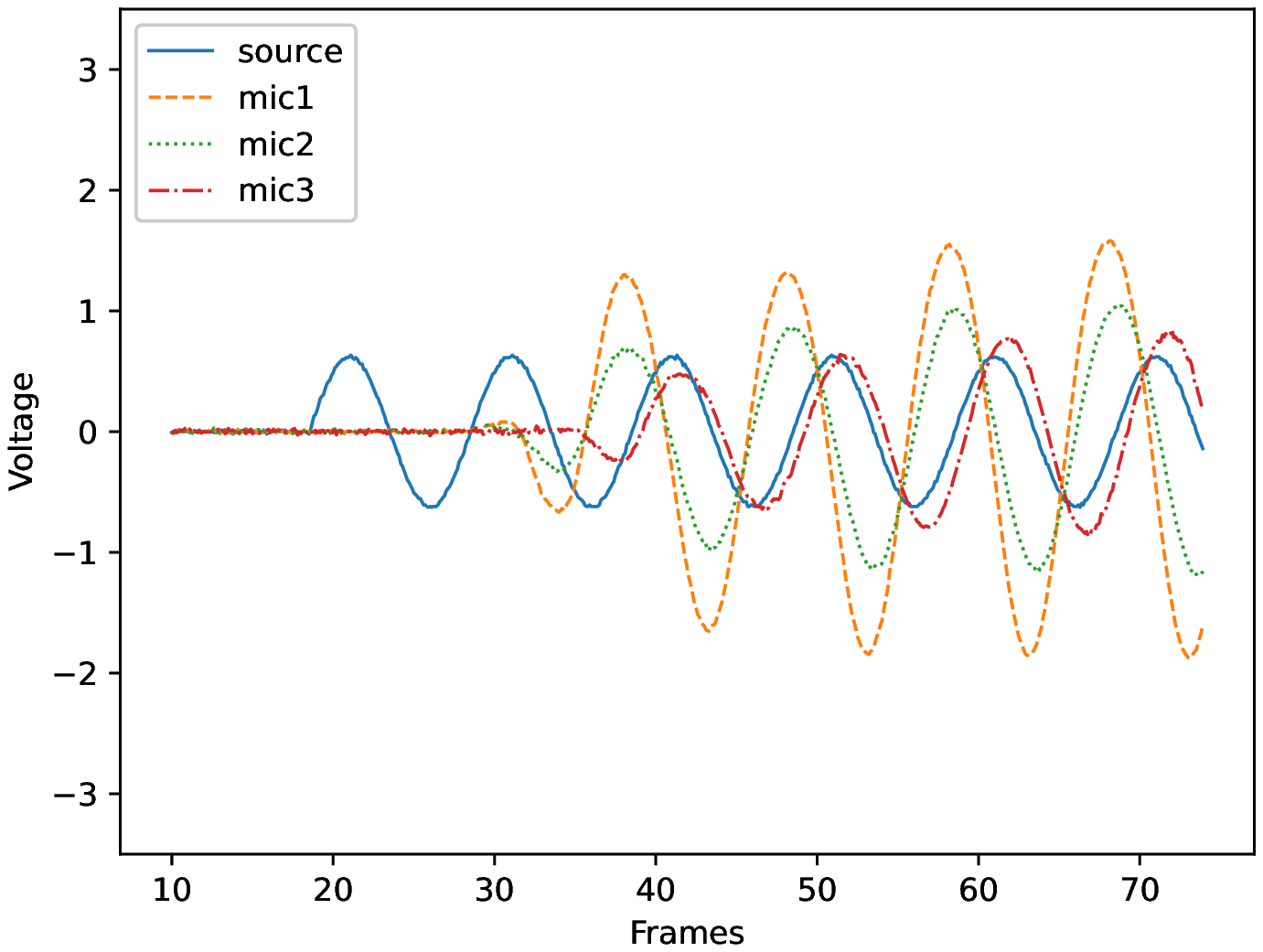}
    \includegraphics[width=0.26\textwidth]{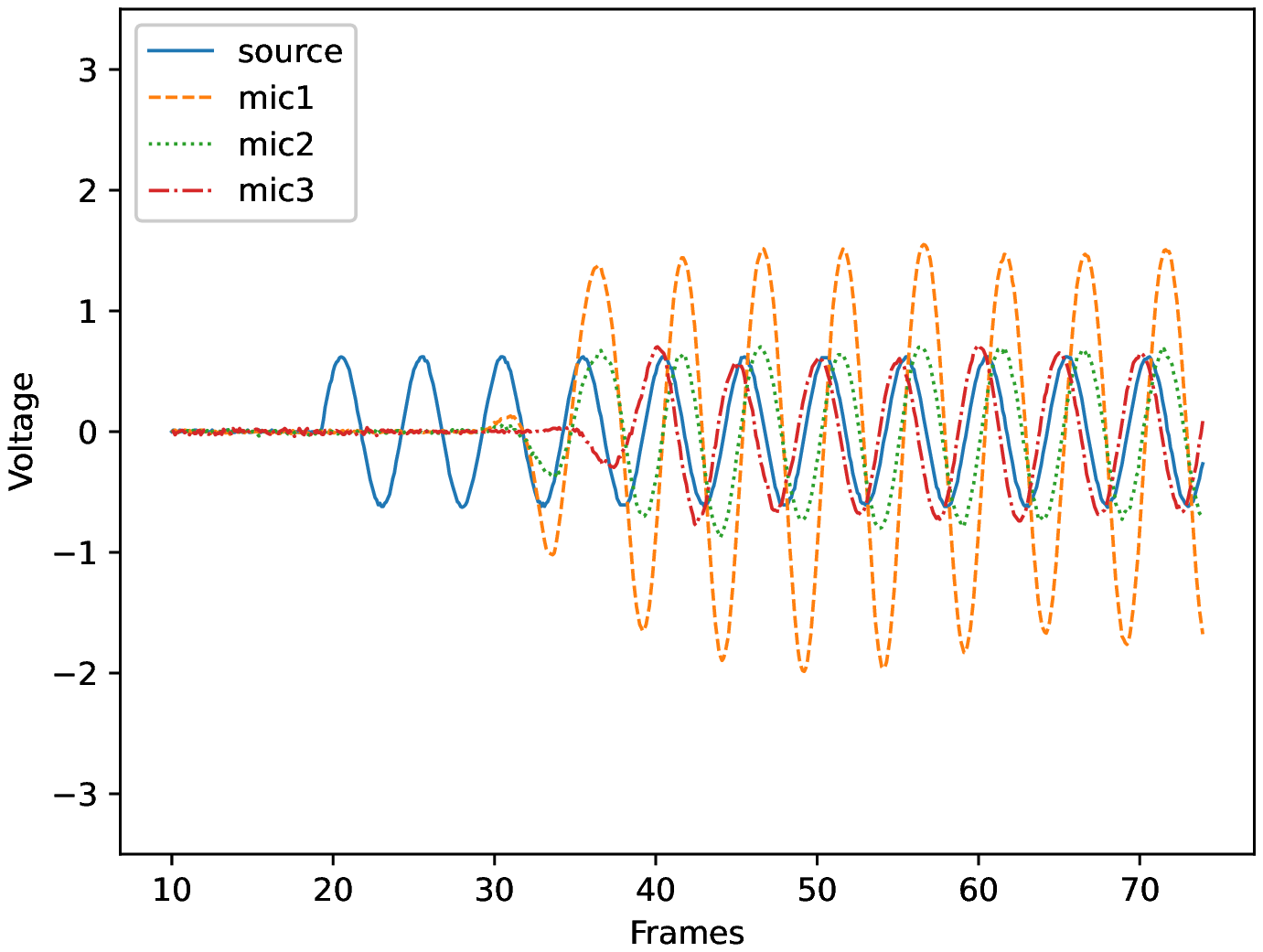}\\
 200Hz  500Hz   1000Hz   2000Hz \\
(b) Ground truth of displacement obtained by microphone.
  \end{center}
  \caption{Visualization of sound wave. Tissue papers are fixed to the floor
  with tapes, and a sound speaker is placed on the right of the images. 
  The sound wave that propagates though the air moves the papers 
from the right to the left, the motion is observed
  accordingly 
and matched with the ground truth obtained by microphones.} 
  \label{fig:floortissue}
\end{figure}
\fi

\begin{figure}[t]
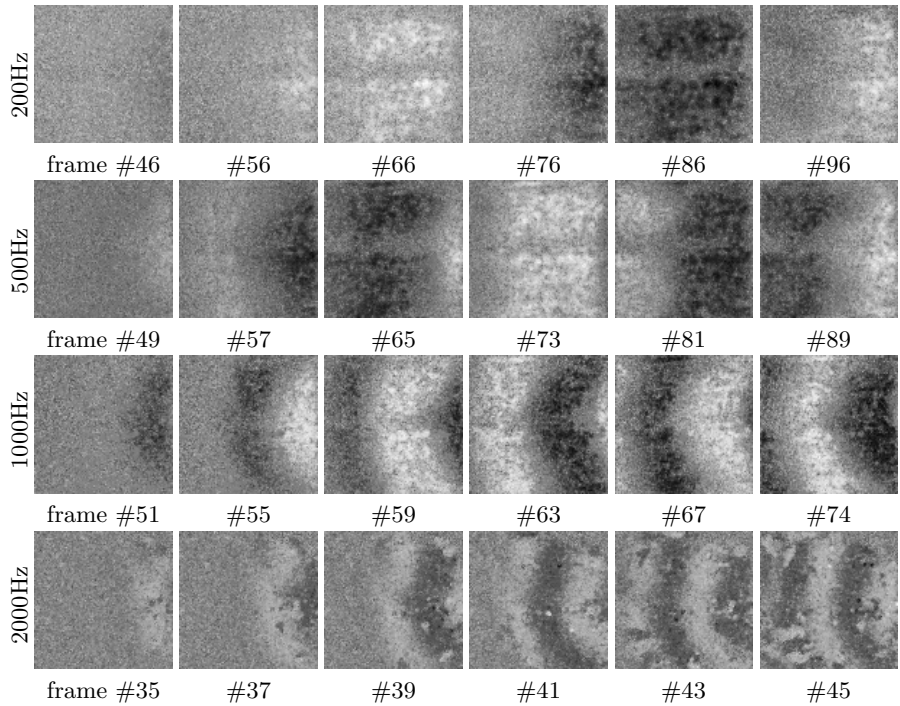

    \begin{center}
      \begin{tabular}{ccccccc}
      \rotatebox{90}{\quad 200Hz}
    &
        \includegraphics[width=0.15\textwidth]{figs/sound_wave/0200hz2/d_0046.png} &
        \includegraphics[width=0.15\textwidth]{figs/sound_wave/0200hz2/d_0056.png} &
        \includegraphics[width=0.15\textwidth]{figs/sound_wave/0200hz2/d_0066.png} &
        \includegraphics[width=0.15\textwidth]{figs/sound_wave/0200hz2/d_0076.png} &
        \includegraphics[width=0.15\textwidth]{figs/sound_wave/0200hz2/d_0086.png} &
        \includegraphics[width=0.15\textwidth]{figs/sound_wave/0200hz2/d_0096.png} \\
        & frame \#46 & \#56 & \#66 & \#76 & \#86 & \#96 \\
    \rotatebox{90}{\quad 500Hz}
    &
        \includegraphics[width=0.15\textwidth]{figs/sound_wave/0500hz2/d_0049.png} &
        \includegraphics[width=0.15\textwidth]{figs/sound_wave/0500hz2/d_0057.png} &
        \includegraphics[width=0.15\textwidth]{figs/sound_wave/0500hz2/d_0065.png} &
        \includegraphics[width=0.15\textwidth]{figs/sound_wave/0500hz2/d_0073.png} &
        \includegraphics[width=0.15\textwidth]{figs/sound_wave/0500hz2/d_0081.png} &
        \includegraphics[width=0.15\textwidth]{figs/sound_wave/0500hz2/d_0089.png} \\
        & frame \#49 & \#57 & \#65 & \#73 & \#81 & \#89 \\
    \rotatebox{90}{\quad 1000Hz}
    &
        \includegraphics[width=0.15\textwidth]{figs/sound_wave/1000hz2/d_0051.png} &
        \includegraphics[width=0.15\textwidth]{figs/sound_wave/1000hz2/d_0055.png} &
        \includegraphics[width=0.15\textwidth]{figs/sound_wave/1000hz2/d_0059.png} &
        \includegraphics[width=0.15\textwidth]{figs/sound_wave/1000hz2/d_0063.png} &
        \includegraphics[width=0.15\textwidth]{figs/sound_wave/1000hz2/d_0067.png} &
        \includegraphics[width=0.15\textwidth]{figs/sound_wave/1000hz2/d_0071.png} \\
        & frame \#51 & \#55 & \#59 & \#63 & \#67 & \#74 \\
    \rotatebox{90}{\quad 2000Hz}
    &
        \includegraphics[width=0.15\textwidth]{figs/sound_wave/2000hz2/d_0035.png} &
        \includegraphics[width=0.15\textwidth]{figs/sound_wave/2000hz2/d_0037.png} &
        \includegraphics[width=0.15\textwidth]{figs/sound_wave/2000hz2/d_0039.png} &
        \includegraphics[width=0.15\textwidth]{figs/sound_wave/2000hz2/d_0041.png} &
        \includegraphics[width=0.15\textwidth]{figs/sound_wave/2000hz2/d_0043.png} &
        \includegraphics[width=0.15\textwidth]{figs/sound_wave/2000hz2/d_0045.png} \\
        & frame \#35 & \#37 & \#39 & \#41 & \#43 & \#45 \\
      \end{tabular}
    \end{center}
    \caption{Visualization of sound wave. Tissue papers are fixed to the floor
    with tapes, and a sound speaker is placed on the right of the images. 
    The sound wave that propagates though the air moves the papers 
  from the right to the left.}
    \label{fig:floortissue}
  \end{figure}

Finally, to visualize a sound wave, 
we set up a speaker and placed tissue papers
that vibrates with the sound which propagates through the air. 
The tissue papers are fixed to the floor with tapes as shown 
in the setup image in \figref{wall_exp}(b) and captured by the high-speed camera with 10,000fps.
We also put three microphones to obtain ground truth data.
The distances from the speaker to the microphones are 34cm, 32cm and 46cm, respectively.
Since the floor is hard enough, the tissue papers' motion is 
considered to be caused only by the air. 
The observed displacements are shown in 
\figref{floortissue}.
The wave-front propagates from the right to the left in wide images.

\section{Conclusion}

This paper proposed a method to observe the minute movement of
objects by using the speckle that is generated by the illumination
of laser light. Since the speckle varies sensitively to the out-of-plane 
movement, it enables to detect the minute movement of objects.
Calculating the displacement without complex optical setup and
calibration is realized based on the approach of embedding
speckle pattern to low-dimensional space.
The embedded vectors, which is calculated for each pixel independently,
are made spatially consistent by estimating transformation matrices.
The displacement is calculated by the consistent embedded vectors.
In the experiments, the calculated displacement is compared
to the movement measured by a micro-stage and accelerometers.
Although the propose method cannot determine the scale of the movement,
it can discriminate the displacement in micrometer accuracy.
The proposed method is applied to various movement and materials,
and succeeded to observe the minute movement of the objects.
In future work, we plan to extend the proposed method to
measure the real scale of the movement with simple and easy calibration.

\clearpage
%
%
\bibliographystyle{splncs}
\bibliography{accv20_speckle}
\end{document}